\documentclass[sigconf]{acmart}
\renewcommand\footnotetextcopyrightpermission[1]{} 
\usepackage{graphicx}
\usepackage[export]{adjustbox}
\usepackage{multirow}
\usepackage{pifont}
\usepackage{enumitem}
\usepackage{balance}
\usepackage{bigfoot}
\usepackage{hyperref}
\usepackage{xcolor}

\DeclareNewFootnote{AAffil}[arabic]
\DeclareNewFootnote{ANote}[fnsymbol]

\AtBeginDocument{%
  \providecommand\BibTeX{{%
    \normalfont B\kern-0.5em{\scshape i\kern-0.25em b}\kern-0.8em\TeX}}}

\setcopyright{acmcopyright}
\copyrightyear{2023}
\acmYear{2023}

\acmConference[MSR-TR-2023-32]{Make sure to enter the correct
  conference title from your rights confirmation email}{September}{2023}

\hypersetup{colorlinks=true,linkcolor=blue,urlcolor=blue}
\begin{document}

\title{Using Large Language Models to Generate, \\Validate, and Apply User Intent Taxonomies}

\makeatletter
\def\thanksAAffil#1#2{
    \protected@xdef\@thanks{\@thanks
        \protect\footnotetextAAffil[#1]{#2}}%
}
\def\thanksANote#1#2{
    \protected@xdef\@thanks{\@thanks
        \protect\footnotetextANote[#1]{#2}}%
}
\makeatother

\author{%
        Chirag Shah\footnotemarkAAffil[1]$^\dagger$$^\ddagger$, 
        Ryen W. White\footnotemarkAAffil[2]$^\dagger$,
        Reid Andersen\footnotemarkAAffil[2],
        Georg Buscher\footnotemarkAAffil[2],
        Scott Counts\footnotemarkAAffil[2],
        Sarkar Snigdha Sarathi Das\footnotemarkAAffil[3]$^\ddagger$,
        Ali Montazer\footnotemarkAAffil[4]$^\ddagger$,
        Sathish Manivannan\footnotemarkAAffil[2],
        Jennifer Neville\footnotemarkAAffil[2],
        Xiaochuan Ni\footnotemarkAAffil[2],
        Nagu Rangan\footnotemarkAAffil[2],
        Tara Safavi\footnotemarkAAffil[2],
        Siddharth Suri\footnotemarkAAffil[2],
        Mengting Wan\footnotemarkAAffil[2],
        Leijie Wang\footnotemarkAAffil[1]$^\ddagger$,
        Longqi Yang\footnotemarkAAffil[2]
}

\affiliation{\vspace*{1em}$^1$University of Washington, $^2$Microsoft, $^3$Pennsylvania State University, $^4$University of Massachusetts Amherst\\ 
$\dagger$Corresponding authors: chirags@uw.edu, ryenw@microsoft.com\\
$\ddagger$Work done while working at Microsoft
\country{USA}
}

\renewcommand{\shortauthors}{Shah et al.}

\begin{abstract}
Log data can reveal valuable information about how users interact with Web search services, what they want, and how satisfied they are. However, analyzing user intents in log data is not easy, especially for emerging forms of Web search such as AI-driven chat. To understand user intents from log data, we need a way to label them with meaningful categories that capture their diversity and dynamics. Existing methods rely on manual or machine-learned labeling, which are either expensive or inflexible for large and dynamic datasets. We propose a novel solution using large language models (LLMs), which can generate rich and relevant concepts, descriptions, and examples for user intents. However, using LLMs to generate a user intent taxonomy and apply it for log analysis can be problematic for two main reasons: (1) such a taxonomy is not externally validated; and (2) there may be an undesirable feedback loop. To address this, we propose a new methodology with human experts and assessors to verify the quality of the LLM-generated taxonomy. We also present an end-to-end pipeline that uses an LLM with human-in-the-loop to produce, refine, and apply labels for user intent analysis in log data. We demonstrate its effectiveness by uncovering new insights into user intents from search and chat logs from the Microsoft Bing commercial search engine. The proposed work's novelty stems from the method for generating purpose-driven user intent taxonomies with strong validation. This method not only helps remove methodological and practical bottlenecks from intent-focused research, but also provides a new framework for generating, validating, and applying other kinds of taxonomies in a scalable and adaptable way with reasonable human effort.
\end{abstract}



\begin{teaserfigure}
    \centering
  \vspace{1mm}
  \includegraphics[width=0.9\textwidth]{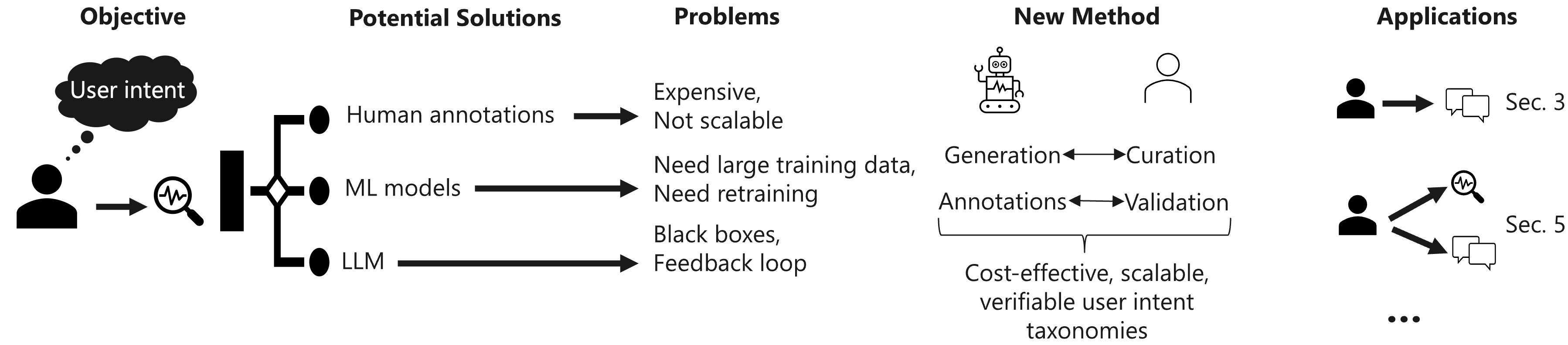}
  \caption{Detecting user intent in search situations is very important, but the potential solutions have different sets of problems. We propose a new method involving a unique collaboration between LLMs and humans to leverage best of both worlds.}
  \label{fig:teaser}
\end{teaserfigure}

\maketitle

\section{Introduction}
\label{ref:intro}
Understanding the purpose or the task behind a user's request in an information access context is highly desired for a search or a recommender system to be able to provide the most relevant and meaningful results \cite{zhang2023efficiently}. However, extracting user intents from log data is extremely difficult due to two main reasons: (1) fluidity in what user intents are or can be; and (2) how these intents can be identified using log data that may not include sufficient context. Additionally, in the case of emerging modalities such as AI-driven chat, users' understanding, usage, and behaviors are rapidly evolving that call for on-demand, task-focused labels and taxonomies. We need new methods to identify, extract, and apply user intents in IR systems, especially those with emerging modalities.

Traditional qualitative methods such as coding and thematic analysis are time-consuming and require human expertise \cite{braun2006using}. Conversely, existing quantitative methods may not capture the nuances and diversity of user intents and experiences \cite{liu2019task}. Large language models (LLMs) have become quite capable of generating coherent texts from various inputs 
\cite{brown2020language}. But can they be useful in reliable and verifiable ways to conduct such intent-focused research?
Specifically, can LLMs help us create and apply a set of labels or categories representing user intents at large-scale to be meaningful for various IR applications? Here, we refer to such set of labels or categories as taxonomy, and acknowledge that a taxonomy does not need to be a single-level or flat hierarchy. Developing and testing a  comprehensive and task-specific taxonomy for user intents can be a first step in various applications, including personalized recommendations, user education, and question/query routing.

There are several attempts in the recent years to use LLMs in various applications ranging from ranking and recommendations to content generation and evaluation (e.g., \cite{benedict2023gen,moore2023empowering,faggioli2023perspectives}). However, many of these works often lack rigor and reliability since LLMs are used as black boxes without a meaningful understanding of their inner workings or there are feedback loops with weak or non-existent validation for the method. Simply focusing on the promising results without sufficient support of scientific rigor in the methodology can lead to misleading and even dangerous outcomes. We believe that while LLMs have shown great promise for aiding us in various informational tasks, they must be used with responsibility and sufficient validation. This leads us to the following research questions (RQs): (1) Can we use LLMs to reliably generate taxonomies for analyzing user intents in log data? (2) Can an LLM correctly apply a user intent taxonomy to annotate logs? (3) If and when can an LLM perform better than human annotators? Or are there uses of LLMs that go beyond just reducing efforts and increasing efficiency?

To address these RQs, we investigated if/how LLMs can help in creating an end-to-end solution for developing user intent taxonomies from AI chat logs. We use GPT-4\footnote{\url{https://openai.com/gpt-4/}}, 
for most of our experiments since it is a leading LLM and we wanted to test the potential of the most advanced models. In the process, we devised a new methodology (illustrated at a high level in Figure \ref{fig:teaser}) for employing any LLM as a collaborator in an iterative qualitative analysis process that leverages its ability to generate summaries, questions, and categories from chat transcripts. 

We already know that LLMs can generate taxonomies, and generate effective annotations \cite{he2023annollm}. Using human-in-the-loop with computational models is also not new. What brings {\bf novelty} in this paper is the methodology that harnesses the potential of an LLM to infer a structure from data and the curation by human assessors in an integrative way. This significantly reduces human effort and provides validity of machine-generated outputs. We demonstrate the value of this by quickly and reliably running the end-to-end pipeline with heterogeneous data from search and chat logs and generating insights. Therein lies the {\bf significance} of this work -- it can remove bottlenecks for research that aims to identify and apply user intents. Beyond that, this methodology can also be useful for generating and applying other kinds of taxonomies in IR and beyond. It will directly benefit researchers and practitioners who develop and evaluate information access systems by providing a streamlined way to understand their users' intentions and adapt systems to them. The {\bf technical soundness} of our work comes from rigorous validation of the proposed method and a set of experiments involving multiple datasets, human assessors, and three different LLMs. Supporting material is available online\footnote{\url{https://anonymous.4open.science/r/LLM-for-taxonomy/}} improving replicability and enabling a wider application of our methods.

\section{Related Work}
\label{sec:related}
We are not the first ones to study user intents, build taxonomies, or use such taxonomies to generate insights from logs. Therefore, before we dive into our novel contributions, viz., using LLMs for generating and using user intents taxonomies with scientific rigor, it is important to briefly review some of the related work.

\subsection{Taxonomy Generation, Validation, and Use}
Taxonomies are hierarchical classifications of concepts, terms, or entities. They can facilitate information seeking, retrieval, or behavior by providing structure, organization, and navigation for users and systems \cite{chakrabarti1997using,yang2012constructing,carrion2019taxonomy}. However, generating and validating taxonomies is a challenging task that requires balancing multiple criteria such as coverage, coherence, consistency, granularity, usability, and adaptability, e.g.,  \cite{kaplan2022introducing,lippell2022taxonomies,raad2015survey,spangler2002interactive}. Moreover, different domains and contexts may have different requirements and preferences for taxonomy design and evaluation \cite{kundisch2021update}. Taxonomies can be generated manually through an iterative process and the research community has developed tools to generate taxonomies automatically from document collections using methods such as clustering \cite{zamir1998web}. We are the first to leverage the power of LLMs to automatically generate taxonomies in a search context, focused specifically on conversational search; we also validate that LLM-based methodology with human assessors. More importantly, we provide a method for other researchers to do the same for their specific needs.

\subsection{Use of LLMs in Research}
The emergence of LLMs has unlocked many opportunities for rapid research advances. LLMs have been used to enable scientific discovery \cite{hope2023computational}, with remarkable progress in areas such as medicine \cite{singhal2023large} and finance \cite{araci2019finbert}. Early language models, such as BERT, and various natural language processing methods have been used to auto-code qualitative data \cite{abram2020methods,grandeit2020using}, although not at a near-human level. LLM-driven AI is capable of qualitative analysis and can generate nuanced results comparable to human researchers \cite{byun2023dispensing}, the costs and benefits of which have been discussed in the literature \cite{bano2023exploring,watkins2023guidance,dwivedi2023so}. 
In information retrieval (IR), LLMs have been shown to be effective in supporting humans in judging document relevance, an activity that is central to search engine design and evaluation \cite{faggioli2023perspectives}. They have also been recently used for synthetic dataset generation to support IR research \cite{jeronymo2023inpars} and richer user modeling to support IR experimentation \cite{li2022user}. Other applications of LLMs in IR have also been discussed at length by others in the IR community \cite{ai2023information}. In this paper, we will show that working directly with humans, LLMs have the potential to support two other critical activities in search engine research and development: intent understanding and generating intent taxonomies. 

\subsection{Log Analysis and Insights}
Search log analysis has been used extensively to gain insights about search interactions, including queries, search engine result page (SERP) clicks, and post-SERP interactions \cite{teevan2007information,craswell2008experimental,white2007investigating}. Analyzing log data has historically been a highly interactive process: researchers first write scripts to extract data, then analyze that data manually using data science tools and methods, and (optionally) human annotators label data to better understand patterns and trends and generate training data for machine learned (ML) models, e.g., \cite{agichtein2012search}. 

\begin{figure*}[ht]
    \centering
    \includegraphics[width=1.8\columnwidth, keepaspectratio]{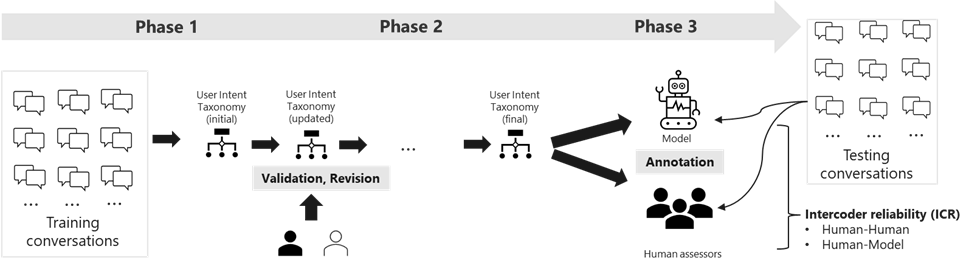}
    \caption{Three phases of user intent taxonomy generation, validation, and application.}
    \label{fig:phases}
\end{figure*}

More recently, there has been an increasing focus on user engagement with conversational search systems \cite{gao2020recent}. Researchers and practitioners have sought to understand user intents and behaviors in the context of chat-based systems \cite{radlinski2017theoretical,vtyurina2020mixed}. Chat is becoming an increasingly popular modality for information seeking, especially in domains where users have complex or exploratory queries, need guidance or clarification, or prefer a conversational style of interaction. Datasets of chat logs have also been created and released to the community to promote research in this emerging area \cite{ren2021wizard}. 

ML models can help support researchers in performing data analysis \cite{heer2019agency}. Recently, from applications in other domains, we have seen that LLMs may have the potential to play a supportive role in the analysis of text data, providing insights and annotations that expedite experiments and reduce human effort \cite{tornberg2023use}. In this paper, for the first time, we introduce methods for using LLMs collaborating with researchers to derive insights and taxonomies from retrospective log data. 

\subsection{Understanding User Intents in Search}
Intent has been a well explored area in IR and various approaches have been proposed for intent understanding and intent representation \cite{kong2015predicting,zhang2019generic}. Taxonomies of search intent can help systems better understand user intent. Several search intent taxonomies generated from search logs have been proposed, e.g., \cite{broder2002taxonomy, jansen2007determining,kellar2007field,rose2004understanding}. These have been generated iteratively via manual inspection of log data and the intent taxonomies include categories for navigation, information, transactions, browsing, and resource finding. We expect chat interactions to exhibit new intents (e.g., creation) compared to traditional search. Automatic generation of query taxonomies has also been attempted \cite{chuang2003automatic,cheng2006query} using query clustering to derive a taxonomy from existing query data and categorization to assign new queries to the taxonomy. Taxonomies have also been used to represent intent in question-answering \cite{bolotova2022non,chen2012understanding}. \citet{xie2002patterns} derived an empirically based classification of search intents that motivate different search behaviors. \citet{mitsui2016extracting} developed a set of information-seeking intentions based on that classification and studied differences in intentions as a function of the search task. Building on this previous work, we leverage LLMs to generate intent taxonomies with humans-in-the-loop and evaluate the performance of LLMs in assigning search activities to intent categories.

\section{Methodology for Generating and Validating a User Intent Taxonomy}
\label{sec:method}
In this section, we describe a new methodology that we developed and tested for employing LLMs to generate a user intent taxonomy that can be used to generate insights and construct hypotheses from log data, focused for this analysis on AI chat logs. 

Let us begin with a problem scenario. We have access to log data from user interactions with an AI-driven chat system (Bing Chat). This data primarily includes user requests and AI responses in natural language. We can analyze this data in a number of ways, answering questions about what the users are doing (topics, domains). However, if we want to understand their intents, we need a set of labels or a taxonomy of intents. In practice, one looks for relevant literature for an existing taxonomy, but if an appropriate taxonomy does not exist, one needs to create it. This can be done by taking an existing taxonomy and modifying it to fit the data or the task (top-down approach), or by building it fresh using the available data (bottom-up approach). Following this, one needs to validate the taxonomy to ensure it meets several criteria for a good taxonomy. Finally, the newly generated taxonomy could be applied to a specific task to generate the desired insights from the data. 

Given that we were interested in analyzing AI chat logs, a seemingly newer type of modality, we found it to be desirable to adopt a bottom-up approach. In this approach, one typically analyzes available data to generate codes or labels, leading to a classification scheme or a taxonomy. As detailed in \cite{broder2002taxonomy, jansen2007determining,kellar2007field,rose2004understanding}, this process could involve one or more researchers and a considerable effort. We wanted to use an LLM to build such a taxonomy using relevant data and instructions. However, given that we do not have enough knowledge about how an LLM is constructed and how it creates or links various concepts from given data, we needed a way to validate LLM's generation and fine-tune it as needed. For that, we used two researchers with many years of experience in doing qualitative analysis and building taxonomies. These researchers guided the taxonomy generation process with the LLM and two human assessors who provided intent labels on data samples to validate and evaluate the LLM-generated output. Once the taxonomy was built using training data and validated, it was applied to annotating test data. In short, our methodology uses LLMs as the backbone of taxonomy generation and application with humans in the loop for curation and validation. 

The outline of our methodology is shown in Figure \ref{fig:phases}. Here, we used GPT-4 as the LLM for generating the taxonomy (Phase 1), engaged human assessors to validate that taxonomy (Phase 2), and then employed both GPT-4 and human assessors to apply it (Phase 3). Through the phases of validation and application, we evaluated not only the generated taxonomy (RQ1), but also GPT-4's ability and potential to perform such research-based tasks reasonably and reliably (RQ2). We now present the details. 

\subsection{Data}
We took a random sample of 1,149 conversations from May-June 2023, available through Bing Chat. Each conversation contained one or more turns of user request and AI response. We ensured that these conversations were in English, however, some of the user requests were interspersed with non-English words. We do not believe this impacted any text processing by GPT-4 for our purposes. We used 1,000 conversations for training (building a user intent taxonomy) and set aside the rest for validation and testing. 

\subsection{Evaluating the Taxonomy}
We first start by describing how a taxonomy should be evaluated. This will inform how we generate the taxonomy using GPT-4 (how we provide prompts), how we validate and revise it, as well as how we measure the effectiveness of the taxonomy for developing insights from logs. Using the relevant literature (some of which is summarized in the previous section) concerning taxonomy generation and validation, we consolidated the following criteria, taken from \citet{raad2015survey}, with appropriate modifications.

\begin{itemize}[leftmargin=*]
    \item {\bf Comprehensiveness}: All the data should be reliably classified using this taxonomy. 
    \item {\bf Consistency}: The taxonomy does not include or allow for any contradictions. 
    \item {\bf Clarity}: The taxonomy should communicate the intended meaning of the defined terms. Definitions should be objective and independent of the context.
    \item {\bf Accuracy}: The definitions, descriptions of classes, properties, and individuals in a taxonomy should be correct. 
    \item {\bf Conciseness}: The taxonomy should not include any irrelevant elements with regards to the user intents in AI chat. 
\end{itemize}

\subsection{Phase 1: Taxonomy Generation}
\label{sec:tax_gen}
Considering how the taxonomy should be evaluated, we constructed a detailed prompt for GPT-4 for generating the first version of taxonomy. We made a few design choices here, including the depth of the taxonomy (single level) and the number of categories (4-6). These suited the intended application: high-level intent understanding. We asked the LLM to generate labels, descriptions, and examples for these categories. The full prompt is in the supporting material.

There were variations in how different versions described the same category. For instance, `Learning' had slightly different meaning and definition in each version of the taxonomy, but generally included concepts and examples of understanding and explanation. Two researchers (two of the co-authors) discussed these three versions and decided to create a consolidated version of the taxonomy, which is shown in Table \ref{tab:taxonomy}.

\begin{table*}[htbp]
\caption{{\small A consolidated version of the user intent taxonomy generated by GPT-4. The examples are collected from the three versions that GPT-4 generated. Slight modifications are made in the user intent title and description using those versions. Short phrases in parentheses next to the intent name are added by researchers for clarity.}}
\label{tab:taxonomy}
\resizebox{\textwidth}{!}{%
\begin{tabular}{|l|l|l|}
\hline
User intent &
  \multicolumn{1}{c|}{Description} &
  \multicolumn{1}{c|}{Examples} \\ \hline
\begin{tabular}[c]{@{}l@{}}Information retrieval (looking for \\ factual information that already exists)\end{tabular} &
  \begin{tabular}[c]{@{}l@{}}The user wants to search, query, or find some information,\\ data, or resources about a topic.\end{tabular} &
  \begin{tabular}[c]{@{}l@{}}Find out the airing dates and channels of women's world cup;\\ Search for information about a phone number; Search for corruption and\\ unemployment statistics for a country.\end{tabular} \\ \hline
\begin{tabular}[c]{@{}l@{}}Problem solving (extracting facts or \\ answers by computing something)\end{tabular} &
  \begin{tabular}[c]{@{}l@{}}The user wants to perform a mathematical or logical operation,\\ such as a conversion, a percentage, a formula, or a function.\end{tabular} &
  \begin{tabular}[c]{@{}l@{}}Compare the size of a human to a hydrogen atom and the observable universe;\\ Compare interest rates for savings accounts; Calculate the distance between a\\ point and a line; Convert a message from Chinese to English.\end{tabular} \\ \hline
\begin{tabular}[c]{@{}l@{}}Learning (satisfying curiosity, helping \\ learn a concept or a phenomenon)\end{tabular} &
  \begin{tabular}[c]{@{}l@{}}The user wants to learn, study, or acquire new skills, concepts, or\\ understanding about a subject. This often involves operations of\\ calculations, comparison, and conversion.\end{tabular} &
  \begin{tabular}[c]{@{}l@{}}Learn about different structural systems; Compare GPT-3 and GPT-4 versions;\\ Explain the difference between Newtonian and non-Newtonian flow.\end{tabular} \\ \hline
Content creation &
  The user wants to write or edit a text for a specific purpose or audience. &
  \begin{tabular}[c]{@{}l@{}}Write an introduction about geothermal energy; Modify a poem into different\\ formats; Improve a report and find adverbs and connectors.\end{tabular} \\ \hline
Leisure &
  \begin{tabular}[c]{@{}l@{}}The user wants to chat or interact with the AI or another agent\\ about various topics or play a game with the AI or another agent.\end{tabular} &
  \begin{tabular}[c]{@{}l@{}}Ask about the AI's sexual orientation and name; Listen to a romantic story;\\ Play tennis and flirt with the user.\end{tabular} \\ \hline
\end{tabular}%
}
\end{table*}

\subsection{Phase 2: Taxonomy Validation}
Next, we provided the taxonomy in Table \ref{tab:taxonomy} to two human annotators along with 10 segments of conversation. They coded them independently, after which we compared their labels. They only had exact agreement three of 10 times. We repeated the whole process with a new sample of 10 segments. It improved, but still had a high level of disagreement (60\%). We, therefore, had another round of discussions and deliberations.
A senior researcher guided these discussions among the annotators. The process involved reviewing each example of disagreement with each annotator explaining their reason behind the label and finally agreeing to which label should be the correct one. As such convergence was reached, they documented how to ensure such instances in the future would receive the same label. This resulted in modified instructions and prompts.

More than trying to reach a higher level of agreement, the goal here was to revise the current version of the taxonomy and develop a better understanding of how a reliable taxonomy could be generated that meets the criteria reported earlier and leads to a common and robust comprehension among the annotators. We learned that the annotators were often extrapolating why a user might have tried to do something.  
That led to most divergence among them. For instance, even when all we could interpret from the data that the user asked for factual information (e.g., ``Does the state of Washington have income tax?''), one of the annotators often extended that to `Learning' intent. It is possible that the user was collecting such information as a part a learning task, but without additional context, it may be impossible to determine that. In such cases, it is advisable to not overextend our understanding and mark the intent based on evidence. Thus, we found it was useful to include in the taxonomy not only positive examples, but also negative examples per category to improve overall clarity. The taxonomy was further modified using negative examples for each category, and the prompt for generating a taxonomy was edited to explicitly ask for negative examples (negative examples are not listed here due to space constraints).

Once we had GPT-4 provide such examples and clarify definitions of `Information Retrieval' and `Learning' categories, we achieved a good match with only 20\% disagreements between the two annotators. In addition, the human assessors did not find a need for any intention not covered here. Thus, the validation state of taxonomy generation was completed and we had the final version of user intent taxonomy (see the supporting material referenced earlier).

\subsection{Phase 3: Taxonomy Application and Testing}
We then took a different set of 124 conversations and asked GPT-4 to code them using the modified taxonomy generated from the above process. We also gave the same instructions to two human annotators. These instructions to humans and the prompt to GPT-4 can be found in the supporting material.

For the human annotators, not a single datapoint was labeled `Other'. GPT-4, on the other hand, marked one out of 124 conversations as `Other'. This further demonstrates comprehensiveness of the taxonomy.
We computed inter-coder reliability (ICR) between two human annotators and found Cohen's kappa to be 0.7620, which indicates a substantial level of agreement \cite{cohen1960coefficient}. 

Next, we asked a third annotator to code these 124 conversations.  
When the three annotators disagreed, we took the majority vote. If all three picked a different label, we labeled that case as `Other'. Finally, we computed ICR between GPT-4 labels and those generated by the majority of human annotators. We computed Cohen's kappa to be 0.7212. This also indicates a substantial level of agreement. 

Overall, what we learned is that when a taxonomy is generated by GPT-4 and verified by humans, it leads to a very high amount of agreement for annotation. That speaks to the {\bf validity} of the generated taxonomy. Also, given that GPT-4's own coding achieves a high level of ICR with human annotators shows that GPT-4 can be used with high {\bf reliability} for the annotation task. 

\subsection{Insights About and From Annotations}
Now that we have demonstrated the end-to-end methodology for generating, validating, and using a taxonomy for understanding user intents in chat logs, let us consider what insights we could derive from the 124 conversation segments 
analyzed by annotators and GPT-4. 

\begin{table}[]
\caption{{\small Confusion matrix for user intent annotations between two human annotators. IR=Information Retrieval, PS=Problem Solving, LR=Learning, CR=Content Creation, LS=Leisure.}}
\label{tab:human-icr}
\begin{tabular}{lcccccc}
 & \multicolumn{1}{l}{} & \multicolumn{1}{l}{} & \multicolumn{2}{l}{\textbf{Annotator-2}} & \multicolumn{1}{l}{} & \multicolumn{1}{l}{} \\ \cline{2-7} 
\multicolumn{1}{l|}{} & \multicolumn{1}{c|}{} & \multicolumn{1}{c|}{\textbf{IR}} & \multicolumn{1}{c|}{\textbf{PS}} & \multicolumn{1}{c|}{\textbf{LR}} & \multicolumn{1}{c|}{\textbf{CR}} & \multicolumn{1}{c|}{\textbf{LS}} \\ \cline{2-7} 
\multicolumn{1}{l|}{} & \multicolumn{1}{c|}{\textbf{IR}} & \multicolumn{1}{c|}{\textbf{42}} & \multicolumn{1}{c|}{2} & \multicolumn{1}{c|}{10} & \multicolumn{1}{c|}{0} & \multicolumn{1}{c|}{0} \\ \cline{2-7} 
\multicolumn{1}{l|}{\multirow{2}{*}{\textbf{Annotator-1}}} & \multicolumn{1}{c|}{\textbf{PS}} & \multicolumn{1}{c|}{0} & \multicolumn{1}{c|}{\textbf{8}} & \multicolumn{1}{c|}{0} & \multicolumn{1}{c|}{4} & \multicolumn{1}{c|}{0} \\ \cline{2-7} 
\multicolumn{1}{l|}{} & \multicolumn{1}{c|}{\textbf{LR}} & \multicolumn{1}{c|}{3} & \multicolumn{1}{c|}{0} & \multicolumn{1}{c|}{\textbf{36}} & \multicolumn{1}{c|}{0} & \multicolumn{1}{c|}{0} \\ \cline{2-7} 
\multicolumn{1}{l|}{} & \multicolumn{1}{c|}{\textbf{CR}} & \multicolumn{1}{c|}{0} & \multicolumn{1}{c|}{0} & \multicolumn{1}{c|}{0} & \multicolumn{1}{c|}{\textbf{8}} & \multicolumn{1}{c|}{0} \\ \cline{2-7} 
\multicolumn{1}{l|}{} & \multicolumn{1}{c|}{\textbf{LS}} & \multicolumn{1}{c|}{1} & \multicolumn{1}{c|}{0} & \multicolumn{1}{c|}{0} & \multicolumn{1}{c|}{0} & \multicolumn{1}{c|}{\textbf{9}} \\ \cline{2-7} 
\end{tabular}
\end{table}

Table \ref{tab:human-icr} presents the confusion matrix between the two human annotators. We can see that Information Retrieval (IR) is the largest category, followed by Learning (LR). The greatest number of times the two annotators disagree is for IR and LR categories. This is understandable since LR always contains IR in a search setting, but it may not always be easy to evaluate if an IR process extends enough to qualify as LR. As noted earlier, this was the biggest factor leading to disagreements among the annotators.

\begin{table}[]
\caption{{\small Confusion matrix for user intent annotations between human and GPT-4 assessments. IR=Information Retrieval, PS=Problem Solving, LR=Learning, CR=Content Creation, LS=Leisure, OT=Other.}}
\label{tab:gpt-icr}
\begin{tabular}{lccccccc}
 & \multicolumn{1}{l}{} & \multicolumn{1}{l}{} & \multicolumn{1}{l}{} & \multicolumn{1}{l}{\textbf{GPT-4}} & \multicolumn{1}{l}{} & \multicolumn{1}{l}{} & \multicolumn{1}{l}{} \\ \cline{2-8} 
\multicolumn{1}{l|}{} & \multicolumn{1}{c|}{} & \multicolumn{1}{c|}{\textbf{IR}} & \multicolumn{1}{c|}{\textbf{PS}} & \multicolumn{1}{c|}{\textbf{LR}} & \multicolumn{1}{c|}{\textbf{CR}} & \multicolumn{1}{c|}{\textbf{LS}} & \multicolumn{1}{c|}{\textbf{OT}} \\ \cline{2-8} 
\multicolumn{1}{l|}{} & \multicolumn{1}{c|}{\textbf{IR}} & \multicolumn{1}{c|}{\textbf{46}} & \multicolumn{1}{c|}{1} & \multicolumn{1}{c|}{5} & \multicolumn{1}{c|}{0} & \multicolumn{1}{c|}{0} & \multicolumn{1}{c|}{1} \\ \cline{2-8} 
\multicolumn{1}{l|}{} & \multicolumn{1}{c|}{\textbf{PS}} & \multicolumn{1}{c|}{0} & \multicolumn{1}{c|}{\textbf{8}} & \multicolumn{1}{c|}{2} & \multicolumn{1}{c|}{0} & \multicolumn{1}{c|}{0} & \multicolumn{1}{c|}{0} \\ \cline{2-8} 
\multicolumn{1}{l|}{\textbf{Human}} & \multicolumn{1}{c|}{\textbf{LR}} & \multicolumn{1}{c|}{12} & \multicolumn{1}{c|}{3} & \multicolumn{1}{c|}{\textbf{26}} & \multicolumn{1}{c|}{1} & \multicolumn{1}{c|}{0} & \multicolumn{1}{c|}{0} \\ \cline{2-8} 
\multicolumn{1}{l|}{} & \multicolumn{1}{c|}{\textbf{CR}} & \multicolumn{1}{c|}{0} & \multicolumn{1}{c|}{0} & \multicolumn{1}{c|}{0} & \multicolumn{1}{c|}{\textbf{10}} & \multicolumn{1}{c|}{0} & \multicolumn{1}{c|}{0} \\ \cline{2-8} 
\multicolumn{1}{l|}{} & \multicolumn{1}{c|}{\textbf{LS}} & \multicolumn{1}{c|}{0} & \multicolumn{1}{c|}{0} & \multicolumn{1}{c|}{3} & \multicolumn{1}{c|}{0} & \multicolumn{1}{c|}{\textbf{5}} & \multicolumn{1}{c|}{0} \\ \cline{2-8} 
\multicolumn{1}{l|}{} & \multicolumn{1}{c|}{\textbf{OT}} & \multicolumn{1}{c|}{1} & \multicolumn{1}{c|}{0} & \multicolumn{1}{c|}{0} & \multicolumn{1}{c|}{0} & \multicolumn{1}{c|}{0} & \multicolumn{1}{c|}{\textbf{0}} \\ \cline{2-8} 
\end{tabular}
\end{table}

Table \ref{tab:gpt-icr} presents the confusion matrix between human annotations (after triaging of three annotators' annotations) and those of GPT-4. Once again, we find that IR and LR are the largest categories and also where we see the most disagreements. Specifically, several (12 out of 124) conversation segments that are marked as LR by humans are labeled as IR by GPT-4. To understand who may be better or more appropriate in picking the labels here, we examined these conversations closely. They all include IR components, but the question is: do they go far enough to indicate an LR  task?  

Unfortunately, we do not have the ground truth here since we do not have access to the original user who conducted the conversation. We interviewed the annotators and found that they extended their understanding of what the users were doing in those segments of conversations to what they might want to do with that information beyond the logged interactions. This often led to a conversation segment being marked as an LR instead of an IR. 
GPT-4 here is strictly labeling the data without making further assumptions, which is desirable. But how consistent this LLM is while making such subjective decisions? To test this, we ran the same test data through GPT-4 four more times and measured ICR among the five sets of annotations by the LLM. We found Fleiss' kappa \cite{fleiss1971measuring} to be 0.8516, indicating a very high level of agreement and consistency.
Therefore, we believe that the labels generated by GPT-4 are better than those generated by humans in this case as they are more objectively and consistently assigned without undesirable extrapolation or assumptions that may not be well-founded, addressing RQ3.
Note that doing the same with humans, viz., having them annotate the same data multiple times to see how consistent they are may be not advisable due to stronger {\em memory effects} for humans (compared to a presumably stateless LLM) or feasible due to the cost involved. This, in itself, offers a benefit of using LLMs, that is, an LLM is likely to be consistent through multiple rounds of annotations on unlimited data, without having any adverse effects resulting from repeated tasks, including fatigue. 

\section{Additional Validations Using Open-Source LLMs}
While we selected GPT-4 as the LLM of choice for developing our method due to its state-of-the-art performance, we wanted to make sure that the results are not typical of GPT-4 and can be replicated by other LLMs, especially ones available as open-source. Therefore, we used Mistral \footnote{\url{https://huggingface.co/docs/transformers/main/model\_doc/mistral}} and Hermes \footnote{\url{https://huggingface.co/NousResearch/Nous-Hermes-Llama2-13b}}, both available from Huggingface as open-source and free LLMs, to do similar experiments for taxonomy generation and application.

\subsection{Single Level Taxonomy Generation}
Similar to the process we instituted for GPT-4 during Phase 1 (Figure \ref{fig:phases}), we fed a prompt and the set of training conversations to Mistral and Hermes. With the assumption and a goal of showing consistency of a prompt across different LLMs, we skipped Phase 2. However, we still needed a way to get some assurance that these two LLMs could also reliably generate user intent taxonomies. For this, we used bootstrapping, where about 80\% of the data was randomly sampled from the available data and provided to a given LLM with a prompt to generate a taxonomy. The prompt remained the same -- the one that resulted from Phase 2 as described before. We ran this process 10 times with each of the three LLMs, each time resulting in a slightly different taxonomy. We performed a few minor manual adjustments to each version as needed. For instance, we substituted `Finding' with `Information Retrieval' and `Enjoy' with `Leisure' for category labels.
Table \ref{tab:generation} shows the union of all generated categories through 30 total runs, with each run generating five categories of user intents.

\begin{table}[htbp]
\caption{{\small Bootstrapping experiments showing frequency of different intent categories over 30 total runs, 10 runs for each of the three LLMs. Top 5 categories in each are bolded.}}
\label{tab:generation}
\begin{tabular}{|l|c|c|c|}
\hline
\textbf{Category}                                                                & \multicolumn{1}{l|}{\textbf{GPT-4}} & \multicolumn{1}{l|}{\textbf{Mistral}} & \multicolumn{1}{l|}{\textbf{Hermes}} \\ \hline
\begin{tabular}[c]{@{}l@{}}Information retrieval/\\ seeking/finding\end{tabular} & {\bf 10}                                  & {\bf 9}                                     & {\bf 10}                                   \\ \hline
Problem solving                                                                  & {\bf 9}                                   & {\bf 8}                                     & {\bf 8}                                    \\ \hline
Learning                                                                         & {\bf 8}                                   & {\bf 10}                                    & {\bf 9}                                    \\ \hline
Content creation                                                                 & {\bf 9}                                   & {\bf 8}                                     & {\bf 8}                                    \\ \hline
Leisure/Entertainment                                                            & {\bf 8}                                   & {\bf 10}                                    & {\bf 7}                                    \\ \hline
Ask for advice/opinion                                                           & 3                                   & 2                                     & 4                                    \\ \hline
Chat                                                                             & 3                                   & 1                                     & 2                                    \\ \hline
Verify                                                                           & 0                                   & 2                                     & 2                                    \\ \hline
\end{tabular}
\end{table}

As Table \ref{tab:generation} shows, while a few instances of taxonomies had intents that were different than what is reported in Table \ref{tab:taxonomy}, most coalesced around the same set of five categories. More importantly, those categories were relatively common across each of the three LLMs. This demonstrated that (1) the prompt we validated and optimized for intent generation is robust and generalizable; and (2) the categories of user intents emerging from the available data are quite stable.

\subsection{Multilevel Taxonomy Generation}
While a taxonomy could have any number of levels, we have thus far focused on generating and applying single level or flat taxonomy only. We should emphasize that this is \emph{not} a limitation of the proposed methodology, but a manifestation of why we are generating such taxonomies and what we intend to do from their applications. The next section presents a specific application where our interest in understanding user intents is to explore how high level tasks are directed and done by two different modalities of information access (search and chat). A small, flat taxonomy is sufficient for this. However, other applications may need more levels or granularity.

To understand how well LLMs can do for generating multilevel taxonomies, we ran Phases 1 and 2 in Figure \ref{fig:phases} with a slightly modified prompt. We added instructions to generate a taxonomy with two levels, with each node not having more than five children. This means we could have up to 30 intent categories for a taxonomy with two levels. This is not unprecedented. There are several works in the literature (e.g., \cite{li2008faceted,mitsui2016extracting,liu2019task}) that have defined and used 20 or more use intents through a multilevel structure.

We ran this modified prompt through all three LLMs -- GPT-4, Mistral, and Hermes -- multiple times with boostrapping. We found that level-1 of the taxonomies maintained the consistency close enough to what is shown in Table \ref{tab:generation}. Examining the subcategories generated for level-2, we found there to be more variance among the three LLMs. A similar behavior can be expected if different human annotators were asked to generate up to 30 categories/subcategories. These initial disagreements among human coders or LLMs could be resolved with more data and more iterations. A full description of this process is beyond the scope for this paper, but a few important findings are reported below.

\begin{enumerate}[leftmargin=*]
    \item Consistencies in subcategories can be improved substantially by holding the level-1 categories constant.
    \item Often subcategories are generated by the model to address very specific instances found in the data. For instance, one subcategory that often emerged was `Look for review' under `Information Retrieval'. We examined the data and discovered several requests in which the user was asking for restaurant or movie reviews. This level of granularity may be essential for some applications, whereas it may add noise for others. If it is the latter, one could prune the taxonomy and remove such subcategories.
    \item To generate meaningful subcategories, we found it to be useful to instruct the LLM that a given subcategory must have at least some minimum number of potential examples from the data, otherwise it should be removed or merged with another subcategory. This is similar to the notion of {\em support} in data mining \cite{morzy2000data}.
    \item In general, it seems essential that each new level of a taxonomy will need its own round of prompt optimization and human intervention for doing appropriate edits and pruning. 
\end{enumerate}

\subsection{Taxonomy Application}
We now move to Phase 3 of Figure \ref{fig:phases} for the two open-source models. We executed this phase with Mistral and Hermes by giving them the same prompt and test data fed to GPT-4. We then measured ICR between each pair of GPT-4, Mistral, Hermes, and human annotations as a way to understand how humans and different LLMs agree on their understanding and application of the taxonomy. The results are shown in Table \ref{tab:all-icr}.

\begin{table}[htbp]
\caption{{\small ICR using Cohen's Kappa}.}
\label{tab:all-icr}
\begin{tabular}{|l|c|c|c|c|}
\hline
                 & \multicolumn{1}{l|}{\textbf{Human}} & \multicolumn{1}{l|}{\textbf{GPT-4}} & \multicolumn{1}{l|}{\textbf{Mistral}} & \multicolumn{1}{l|}{\textbf{Hermes}} \\ \hline
\textbf{Human}   & 0.7620                              & --                                  & --                                    & --                                   \\ \hline
\textbf{GPT-4}   & 0.7212                              & --                                  & --                                    & --                                   \\ \hline
\textbf{Mistral} & 0.6943                              & 0.6343                              & --                                    & --                                   \\ \hline
\textbf{Hermes}  & 0.6521                              & 0.5732                              & 0.6772                                & --                                   \\ \hline
\end{tabular}
\end{table}

As we can see, the ICR values as measured by Cohen's Kappa were moderate (0.41-0.60) to substantial (0-61-0.80). This is an indirect validation of the taxonomy that was generated as it indicates that the categories and their descriptions were comprehensive, consistent, clear, accurate, and concise enough that different entities -- humans and three LLMs -- could apply them to new data in a very similar way. There are still other elements of a good taxonomy that cannot be measured by such ICR scores. For example, the usefulness and objectivity of the taxonomy may depend on the application. In the next section, we consider a specific application of user intent taxonomy as a way to further validate our proposed methodology.

\section{Application of user intent pipeline}
\label{sec:pipeline}
\begin{figure*}[ht]
    \centering
    \includegraphics[width=2.0\columnwidth, keepaspectratio]{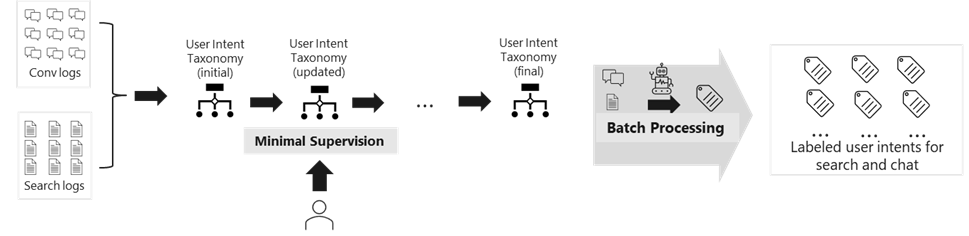}
    \caption{Using LLM in an end-to-end pipeline for generating, validating, and applying a taxonomy for user intents.}
    \label{fig:pipeline}
\end{figure*}

Can we apply the proposed methodology to other, perhaps more challenging tasks that call for identifying user intents? We were driven by a hypothesis that there is a shift in users' behaviors and intents happening with the emergence of AI chat modality. To test this, we built a pipeline using the method proposed here.
Depicted in Figure \ref{fig:pipeline}, the pipeline shows that   
there is still a human in the loop but given that we know how to construct and apply a reliable taxonomy using a process that is already validated, we can now use human intervention in shaping the process and doing light touch validations. 
Notably, the intent taxonomy here is created with heterogeneous log data containing only user requests.

\subsection{Stepwise Process for the Full Pipeline}
In the steps below, we describe this process as a full pipeline for how one could leverage LLM for analyzing log data. Through the process, we will also focus on evaluating various aspects of the generated taxonomy. These aspects include comprehensiveness, consistency, clarity, accuracy, and conciseness.

But why generate a new taxonomy if there are several existing taxonomies, including the one generated in the previous section? While one of these taxonomies could be fitting, it is desirable that we have a taxonomy that is rooted in specific data and application we have under consideration. Given the small cost of generating a taxonomy may also justify at least attempting to construct a new taxonomy and deciding if it is more fitting than anything available.

\noindent
{\bf Step-1: Identify Application and Data}

The first step is to identify what kind of data we want to extract user intents from and why. Here, we are interested in understanding how users have different or overlapping intents between two modalities -- search and chat. Using log data available to use from Bing Search and Bing Chat, we needed to first build a new user intent taxonomy and then apply that taxonomy to annotate log data. We extracted a random sample of users who had used both Bing Search and Bing Chat from May-June 2023. From those users, we extracted 2,456 queries and 15,531 chat requests 
they had sent to the respective services. We used 500 search queries and 500 chat requests (a total of 1,000 user inquiries to Bing) for training and set aside the rest for testing. We randomized their order, forming our training set with 1,000 data points. There are two ways this data is different from the data used  earlier, making this a different and a more challenging task: (1) it contains log data from different modalities; and (2) it does not contain system response.

\noindent
{\bf Step-2: Build/Fine-tune Taxonomy with Human-in-the-loop}

To get started with the LLM (here, GPT-4), we built the initial prompt that explained what we are trying to do, what the data contains, and what are some of the criteria or constraints. For example, we indicated that we are looking for a taxonomy of user intents with no more than five categories and the criteria for a good taxonomy are comprehensiveness, consistency, clarity, accuracy, and  conciseness as defined earlier. The full prompt is given in the supporting material. This resulted in the zero-shot version of the taxonomy with the following five categories: 

\begin{itemize}[leftmargin=*]
    \item {\em Ask for Advice or Recommendation}: The intent to seek suggestions, opinions, or guidance from others on a specific topic or situation. 
    \item {\em Create}: The intent to use AI tools or platforms to generate, edit, or manipulate information objects. 
    \item {\em Information Retrieval}: The intent to find existing information or answers on the internet. 
    \item {\em Learn}: The intent to acquire new knowledge or skills on a subject of interest. 
    \item {\em Leisure}: The intent to enjoy oneself by engaging in amusing activities such as games, jokes, stories, etc. 
\end{itemize}

This is not very different from what we saw in the previous section. 
However, we noticed that the descriptions and examples associated with these labels were different and more suitable for our purpose. 
Even if there was evidence or intuition that an existing taxonomy would be sufficient for our purpose, given the reasonable cost for generating a new taxonomy, it may be desirable to go through these two steps to validate and revise that taxonomy with a goal to fare better along the criteria for a quality taxonomy described before.

\noindent
{\bf Step-3: Measure Taxonomy Comprehensiveness/Consistency}

We now need to test how complete and consistent this taxonomy is. For that, we fed it as a prompt to the LLM and have it label each of the samples we used before separately. This time, we asked it explicitly to label anything that does not fit the provided labels as `Other'. We found that no sample fell under this category. This indicated that the taxonomy was comprehensive and consistent.

\noindent
{\bf Step-4: Improve Taxonomy Clarity}

Next, we asked the LLM to expand each category label with more description and examples to improve its clarity. Taking the lesson from before, we also asked GPT-4 for negative examples per category, improving on the taxonomy's clarity.

\noindent
{\bf Step-5: Measure Validity and Accuracy}

As the final step of validation and refinement, we asked the LLM to use the constructed taxonomy to label the same data that was used to generate the taxonomy. Normally, this is not a practice for testing, but here we are looking for internal validity and accuracy of the taxonomy. Recall that we had 1,000 data points (500 search queries and 500 chat requests) for training. Once the LLM labeled each of these, we took a random sample of 100 and manually checked if the assigned label follows the definition for that label as generated before. We found that the answer was `yes' for 95 of these samples and that there was no sample assigned `Other' category. This analysis provided us with the assurance the taxonomy was valid and accurate. 

\noindent
{\bf Step-6: Perform Annotations and Measure Conciseness}

Finally, we ran our test data -- 1,956 search queries and 15,031 chat requests -- through GPT-4 with the final version of the taxonomy as a part of the prompt. This prompt is given in the supporting material. We found that no sample was marked with `Other' label, ensuring that all the important concepts were covered. Also, no category had too few (subjective, but in our case $<2\%$) samples, indicating that the taxonomy was concise. 

\subsection{Insights about Intents in Search vs. Chat}
The steps above demonstrated that we could create a user intent taxonomy fulfilling all the criteria for a high-quality, reliable, and robust taxonomy. 
If one needs additional assurance for annotation quality, at this point, a small sample of this test data can be taken for human assessment and ICR can be computed between that assessment and the one from the LLM. Such data samples can also be used as a way to test for bias in the taxonomy that may have resulted from any biases present in the underlying LLM.

For our purposes, we decided to move on to deriving insights from this test data. Given that we had an uneven number of queries and chat requests, we normalized them around each intent category by counting the number of instances for a given modality w.r.t. an intent and dividing it by the total instances of search and chat for that intent.

\begin{figure}[ht]
    \centering
    \includegraphics[width=0.85\columnwidth, keepaspectratio]{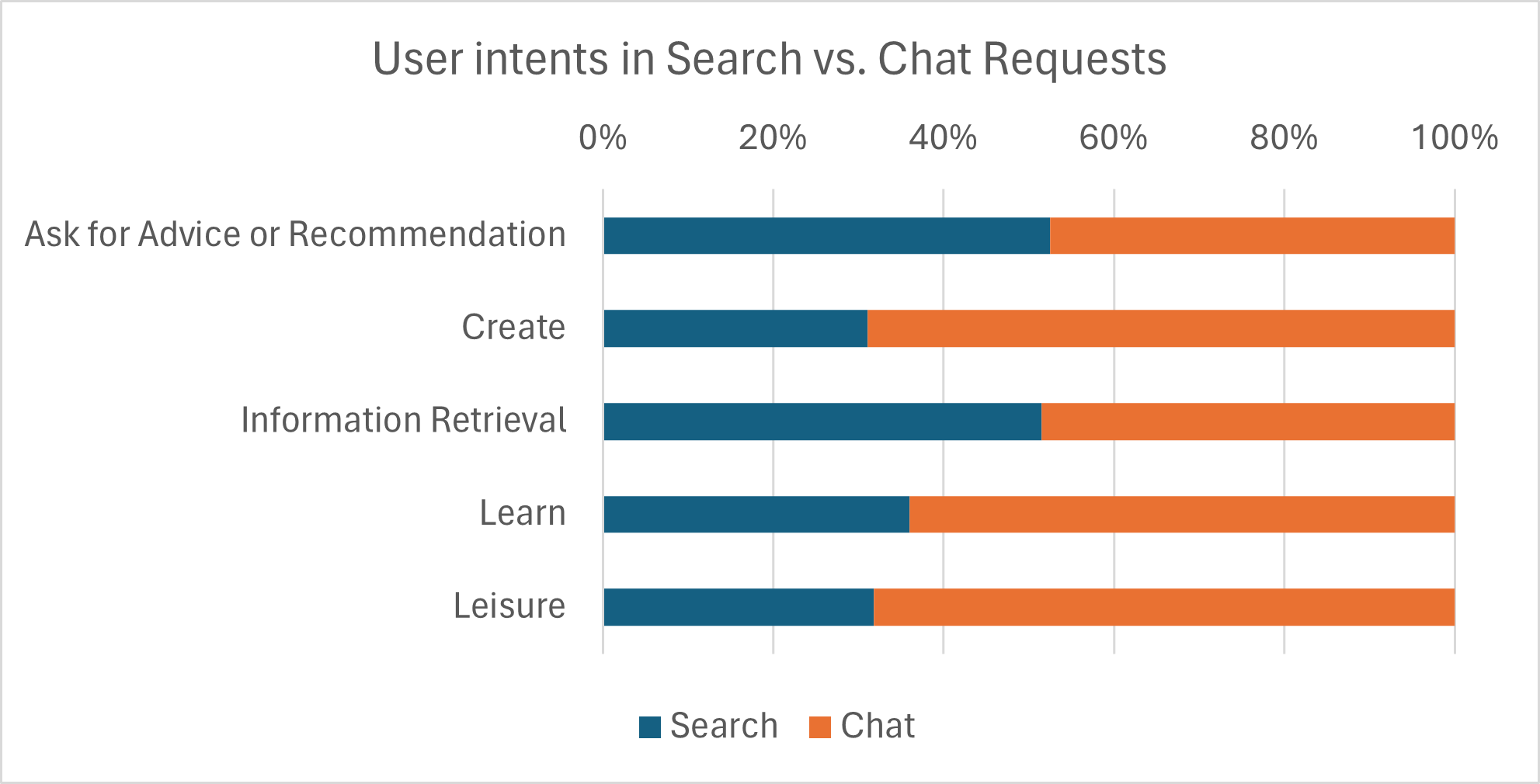}
    \caption{Comparing user intents between search and chat.}
    \label{fig:search_chat}
\end{figure}

Figure \ref{fig:search_chat} shows the distribution of user intents for search and chat. As shown, `Ask for Advice or Recommendation' and `Information Retrieval' are almost evenly distributed between search and chat, with a little bit of skew toward search. The other three categories (`Create', `Learn', and `Leisure') are heavily leaning toward chat. This requires a close examination.

First off, it is important to understand that this figure shows a view from the user intent perspective. If a user had an intent or a task related to one of the five intents considered here, where would they go -- search or chat? We found that while they could use either for their `Information Retrieval' or `Ask for Advice or Recommendation' needs, they are favoring (with a significant tilt toward) chat for their create, learn, and leisure intents. `Create' makes sense because it is more suitable to use a generative AI tool like Bing Chat for a creation task than a search engine. Of course, users are still sending search engines with create-related requests, but we hypothesize that as generative IR tools such as AI chat become more capable and known, that intent will shift more dramatically from search to chat. Similarly, `Leisure' makes sense because here the user intends to have a social or open-ended conversation, which is by design supported through chat. Perhaps more interesting finding here is with respect to `Learn'. Learning is considered to be a higher-level goal or task in information seeking \cite{vakkari2016searching}. While people have used keyword-based search system for such a task, with chat-based generative IR systems, the intent fits the modality more appropriately. Through a manual inspection of some of the logs available to us, we could see that users are indeed issuing higher-level and complex requests, often associated with learning, through the chat interface. We should note this with a caveat that our unit of analysis here is a single request from the user. It is possible that the user issued multiple queries in a given search session to accomplish their learning task. Even then, it is interesting to learn that users are preferring to issue their single-request learning requests through chat. As the information access systems with emerging technology such as generative AI and conversation-based modalities chart their course to support users in new meaningful ways, they should consider their designs from a user intent perspective.

\subsection{Steps for Generating Intent Taxonomies}
\label{sec:steps}
Now that we have described the methodology and demonstrated how it could be executed using an application, we summarize the lessons from these experiments and provide a guidance to anyone who wants to use LLMs for generating, validating, or applying user intent taxonomies. 

\begin{enumerate}[leftmargin=*]
\item {\em Identify Application and Data}. A taxonomy must fulfill the purpose for which it is built. That also means an existing taxonomy may not be right for your application. Assuming you want to build a purpose-driven taxonomy, prepare a detailed description of what user intent means for your application and how it should be used. For instance, in our case, it was important for us to stay focused on users' actions in a task rather than the objects when recognizing intents. This means we would not want intents that are tied to an object (e.g., `finding information about tax') and stay close to a general action or objective (e.g., `information retrieval'). It is also important to have as clean data as possible for an LLM to process it appropriately. Depending on which LLM you use, you may need to check for input requirements such as the size and language of input tokens. 

\item {\em Build and Fine-tune Taxonomy}. Pay attention to the first prompt you prepare for building the taxonomy. Add details of your application/task, your criteria for a good taxonomy, and relevant constraints (e.g., number of levels, number of categories, length of a label). We recommend bootstrapping to build different versions of the taxonomy to see how sensitive it is to the data used, like what we showed in Section \ref{sec:tax_gen}. 

\begin{enumerate}[leftmargin=*]
    \item {\em Check for Comprehensiveness}. Construct a prompt to annotate input data using the taxonomy built. Feed the training data to the LLM with this prompt to have it label the data. If what falls under `Other' category is more than, say, 5\% of your data (which could vary depending on the specific application and data characteristics), you may need to create additional categories or levels to make your taxonomy more complete.
    \item {\em Check for Consistency}. Assuming your training data is of a reasonable size, it may not be feasible to manually check labels for each of the samples, but you can take an appropriate random sample and see if the LLM consistently applied the definitions of various categories. You could also perform multiple runs of Step-2 and see if a sample gets labeled the same way every time. Since this is an iterative and exploratory process, you can decide how far and deep you want to go before seeing good enough convergence and consistency.
    \item {\em Check for Bias}. There are many ways to detect bias in the LLM output \cite{mokander2023auditing}. For example, we could test with multiple LLMs and compare the results to see if the current model is an outlier. We could have humans evaluate the taxonomies for unexpected intent distributions (e.g., significant skew toward intent categories that are more popular than we would expect to see). We could statistically compare of intent distributions from the LLM with intent distributions from human-labeled data (we would expect them to be quite similar). If bias is detected through any of these measures, we could adjust the prompts, use a less-biased LLM, or even decide not to use this approach.
    \item {\em Improve Taxonomy's Clarity}. Once the above steps are done reasonably well or skipped as appropriate, your taxonomy is now fixed. At this point, you may ask the LLM to revise and expand the definition or description for each of your labels to improve its clarity. Often, feeding appropriate examples (positive and negative) can be useful -- similar to how a human annotator is trained. 
\end{enumerate}

\item {\em Measure Accuracy and Conciseness}. As a final and another optional step, you can give the LLM test data, ensuring this data was not used before for any training purposes, for doing annotations using the final version of the taxonomy. Is there any category that does not get enough samples? If so, you may decide to remove that category to improve your taxonomy's conciseness. Note that if you do this, you may have to repeat some of the steps from before because now those samples will fall under other categories, which may affect some of the criteria evaluated before. Now take a random sample of labeled data and have a human annotator label it using the same instructions given to the LLM as prompt. Measure the ICR between human annotations and those from the LLM for the same data. This measurement will give you a sense of how accurate or valid your taxonomy is as well as your LLM's annotation capabilities. If at this point you have taken all the steps before (or skipped them as appropriate) and found a high enough ICR, your taxonomy and your LLM have been thoroughly tested. 
\end{enumerate}

\section{Conclusion}
Identifying user intents in online information access is highly crucial for most search and recommender systems. But doing so is often very challenging. Even if one has a pre-defined taxonomy of user intents, training an ML model or using such a model to annotate rapidly changing behavioral traits in new modalities such as AI chat can be expensive or infeasible. LLMs are shown to be effective at extracting concepts, descriptions or summaries, and examples from given set of text. This could be used for building and using taxonomies containing user intents, but there is a danger of creating a feedback loop without a clear evaluation. 

In this paper we presented a novel methodology for using LLMs in generating, validating, and using taxonomies for identifying user intents in various applications. The methodology was demonstrated using an application of understanding user intents in AI chat logs. A case study was then presented with the application of contrasting user intents between search and chat. The results from both the applications are intriguing, presenting a set of new hypotheses and calling for further explorations. However, the primary contribution of this paper is the methodology for deploying LLMs in such research tasks.

As a reference from our own experiments, building the full pipeline in that case study (Section \ref{sec:pipeline}) took less than half the time and effort compared to the process executed for developing the method (Section \ref{sec:method}). The process described in Section \ref{sec:steps} will take substantially less even if all the optional steps are executed. Such efficiency is more than simply reducing the effort for one set of experiments. Emerging technologies such as AI-driven chat are being discovered and used by a large set of new users. As they become more accustomed, we can expect to see shifts in the kind of tasks they do and the kinds of intents they have with these modalities. The approach presented here will allow researchers to adapt to these evolving intents quickly and at lower cost and effort.  

Through the development of this methodology, we learned that we could use various LLMs for a zero-shot construction of a user intent taxonomy, given some log data with user requests in natural language. While this taxonomy was of a reasonable quality, we found the need to have human verification and fine tuning to ensure that such a taxonomy meets various criteria commonly expected in the literature and in practice, including comprehensiveness, consistency, clarity, accuracy, and conciseness (as well as additional criteria that may be especially important for AI-driven chat systems, such as adaptability and scalability to new patterns of interaction and diverse user contexts). Through the development of this methodology and its subsequent application in a different case study, we showed how these criteria can be reliably fulfilled using an LLM and human-in-the-loop.

In fact, the experiments described here are not the only recent examples of the proposed method being applied to IR applications. For examples, \citet {amirizaniani2024developing, amirizaniani2024auditllm} recently used this method to audit LLMs using a multi-probe approach and human-in-the-loop. We need to experiment with different taxonomy designs and different applications that require more levels or granularity to understand the flexibility and scalability of the proposed methodology.

In this regard, we conclude that an LLM can serve as a collaborator or a copilot rather than a replacement for human researchers. This human-LLM collaboration can yield not only faster construction and validation of a new user intent taxonomy, but also higher quality outputs with crisply defined labels, descriptions, and examples. Once the phases of construction and validation are done, the LLM can very effectively and accurately perform the annotation task, turning from copilot to autopilot. This can allow us to analyze large-scale data and generate insights automatically. Finally, we found that often LLMs not only made things go faster, but also better. In cases of disagreements with human annotators, we found that GPT-4 was producing user intent labels truer to the data given rather than extrapolating to situations for which we lacked evidence. In short, the work reported here charts a new territory for using LLMs as collaborators and consignors 
for user intent analysis in an effective, efficient, and responsible manner. 

\newpage

\bibliographystyle{ACM-Reference-Format}
\bibliography{references}


\begin{thebibliography}{59}


\ifx \showCODEN    \undefined \def \showCODEN     #1{\unskip}     \fi
\ifx \showDOI      \undefined \def \showDOI       #1{#1}\fi
\ifx \showISBNx    \undefined \def \showISBNx     #1{\unskip}     \fi
\ifx \showISBNxiii \undefined \def \showISBNxiii  #1{\unskip}     \fi
\ifx \showISSN     \undefined \def \showISSN      #1{\unskip}     \fi
\ifx \showLCCN     \undefined \def \showLCCN      #1{\unskip}     \fi
\ifx \shownote     \undefined \def \shownote      #1{#1}          \fi
\ifx \showarticletitle \undefined \def \showarticletitle #1{#1}   \fi
\ifx \showURL      \undefined \def \showURL       {\relax}        \fi
\providecommand\bibfield[2]{#2}
\providecommand\bibinfo[2]{#2}
\providecommand\natexlab[1]{#1}
\providecommand\showeprint[2][]{arXiv:#2}

\bibitem[Abram et~al\mbox{.}(2020)]%
        {abram2020methods}
\bibfield{author}{\bibinfo{person}{Marissa~D Abram}, \bibinfo{person}{Karen~T
  Mancini}, {and} \bibinfo{person}{R~David Parker}.}
  \bibinfo{year}{2020}\natexlab{}.
\newblock \showarticletitle{Methods to integrate natural language processing
  into qualitative research}.
\newblock \bibinfo{journal}{\emph{International Journal of Qualitative
  Methods}}  \bibinfo{volume}{19} (\bibinfo{year}{2020}),
  \bibinfo{pages}{1609406920984608}.
\newblock


\bibitem[Agichtein et~al\mbox{.}(2012)]%
        {agichtein2012search}
\bibfield{author}{\bibinfo{person}{Eugene Agichtein}, \bibinfo{person}{Ryen~W
  White}, \bibinfo{person}{Susan~T Dumais}, {and} \bibinfo{person}{Paul~N
  Bennet}.} \bibinfo{year}{2012}\natexlab{}.
\newblock \showarticletitle{Search, interrupted: understanding and predicting
  search task continuation}. In \bibinfo{booktitle}{\emph{Proceedings of the
  35th international ACM SIGIR conference on Research and development in
  information retrieval}}. \bibinfo{pages}{315--324}.
\newblock


\bibitem[Ai et~al\mbox{.}(2023)]%
        {ai2023information}
\bibfield{author}{\bibinfo{person}{Qingyao Ai}, \bibinfo{person}{Ting Bai},
  \bibinfo{person}{Zhao Cao}, \bibinfo{person}{Yi Chang},
  \bibinfo{person}{Jiawei Chen}, \bibinfo{person}{Zhumin Chen},
  \bibinfo{person}{Zhiyong Cheng}, \bibinfo{person}{Shoubin Dong},
  \bibinfo{person}{Zhicheng Dou}, \bibinfo{person}{Fuli Feng}, {et~al\mbox{.}}}
  \bibinfo{year}{2023}\natexlab{}.
\newblock \showarticletitle{Information Retrieval Meets Large Language Models:
  A Strategic Report from Chinese IR Community}.
\newblock \bibinfo{journal}{\emph{arXiv preprint arXiv:2307.09751}}
  (\bibinfo{year}{2023}).
\newblock


\bibitem[Amirizaniani et~al\mbox{.}(2024a)]%
        {amirizaniani2024auditllm}
\bibfield{author}{\bibinfo{person}{Maryam Amirizaniani}, \bibinfo{person}{Tanya
  Roosta}, \bibinfo{person}{Aman Chadha}, {and} \bibinfo{person}{Chirag Shah}.}
  \bibinfo{year}{2024}\natexlab{a}.
\newblock \showarticletitle{AuditLLM: A Tool for Auditing Large Language Models
  Using Multiprobe Approach}.
\newblock \bibinfo{journal}{\emph{arXiv preprint arXiv:2402.09334}}
  (\bibinfo{year}{2024}).
\newblock


\bibitem[Amirizaniani et~al\mbox{.}(2024b)]%
        {amirizaniani2024developing}
\bibfield{author}{\bibinfo{person}{Maryam Amirizaniani}, \bibinfo{person}{Jihan
  Yao}, \bibinfo{person}{Adrian Lavergne}, \bibinfo{person}{Elizabeth~Snell
  Okada}, \bibinfo{person}{Aman Chadha}, \bibinfo{person}{Tanya Roosta}, {and}
  \bibinfo{person}{Chirag Shah}.} \bibinfo{year}{2024}\natexlab{b}.
\newblock \showarticletitle{Developing a Framework for Auditing Large Language
  Models Using Human-in-the-Loop}.
\newblock \bibinfo{journal}{\emph{arXiv preprint arXiv:2402.09346}}
  (\bibinfo{year}{2024}).
\newblock


\bibitem[Araci(2019)]%
        {araci2019finbert}
\bibfield{author}{\bibinfo{person}{Dogu Araci}.}
  \bibinfo{year}{2019}\natexlab{}.
\newblock \showarticletitle{Finbert: Financial sentiment analysis with
  pre-trained language models}.
\newblock \bibinfo{journal}{\emph{arXiv preprint arXiv:1908.10063}}
  (\bibinfo{year}{2019}).
\newblock


\bibitem[Bano et~al\mbox{.}(2023)]%
        {bano2023exploring}
\bibfield{author}{\bibinfo{person}{Muneera Bano}, \bibinfo{person}{Didar
  Zowghi}, {and} \bibinfo{person}{Jon Whittle}.}
  \bibinfo{year}{2023}\natexlab{}.
\newblock \showarticletitle{Exploring Qualitative Research Using LLMs}.
\newblock \bibinfo{journal}{\emph{arXiv preprint arXiv:2306.13298}}
  (\bibinfo{year}{2023}).
\newblock


\bibitem[B{\'e}n{\'e}dict et~al\mbox{.}(2023)]%
        {benedict2023gen}
\bibfield{author}{\bibinfo{person}{Garbiel B{\'e}n{\'e}dict},
  \bibinfo{person}{Ruqing Zhang}, {and} \bibinfo{person}{Donald Metzler}.}
  \bibinfo{year}{2023}\natexlab{}.
\newblock \showarticletitle{Gen-IR@ SIGIR 2023: The First Workshop on
  Generative Information Retrieval}. In \bibinfo{booktitle}{\emph{Proceedings
  of the 46th International ACM SIGIR Conference on Research and Development in
  Information Retrieval}}. \bibinfo{pages}{3460--3463}.
\newblock


\bibitem[Bolotova et~al\mbox{.}(2022)]%
        {bolotova2022non}
\bibfield{author}{\bibinfo{person}{Valeriia Bolotova},
  \bibinfo{person}{Vladislav Blinov}, \bibinfo{person}{Falk Scholer},
  \bibinfo{person}{W~Bruce Croft}, {and} \bibinfo{person}{Mark Sanderson}.}
  \bibinfo{year}{2022}\natexlab{}.
\newblock \showarticletitle{A non-factoid question-answering taxonomy}. In
  \bibinfo{booktitle}{\emph{Proceedings of the 45th International ACM SIGIR
  Conference on Research and Development in Information Retrieval}}.
  \bibinfo{pages}{1196--1207}.
\newblock


\bibitem[Braun and Clarke(2006)]%
        {braun2006using}
\bibfield{author}{\bibinfo{person}{Virginia Braun} {and}
  \bibinfo{person}{Victoria Clarke}.} \bibinfo{year}{2006}\natexlab{}.
\newblock \showarticletitle{Using thematic analysis in psychology}.
\newblock \bibinfo{journal}{\emph{Qualitative research in psychology}}
  \bibinfo{volume}{3}, \bibinfo{number}{2} (\bibinfo{year}{2006}),
  \bibinfo{pages}{77--101}.
\newblock


\bibitem[Broder(2002)]%
        {broder2002taxonomy}
\bibfield{author}{\bibinfo{person}{Andrei Broder}.}
  \bibinfo{year}{2002}\natexlab{}.
\newblock \showarticletitle{A taxonomy of web search}. In
  \bibinfo{booktitle}{\emph{ACM Sigir forum}}, Vol.~\bibinfo{volume}{36}. ACM
  New York, NY, USA, \bibinfo{pages}{3--10}.
\newblock


\bibitem[Brown et~al\mbox{.}(2020)]%
        {brown2020language}
\bibfield{author}{\bibinfo{person}{Tom Brown}, \bibinfo{person}{Benjamin Mann},
  \bibinfo{person}{Nick Ryder}, \bibinfo{person}{Melanie Subbiah},
  \bibinfo{person}{Jared~D Kaplan}, \bibinfo{person}{Prafulla Dhariwal},
  \bibinfo{person}{Arvind Neelakantan}, \bibinfo{person}{Pranav Shyam},
  \bibinfo{person}{Girish Sastry}, \bibinfo{person}{Amanda Askell},
  {et~al\mbox{.}}} \bibinfo{year}{2020}\natexlab{}.
\newblock \showarticletitle{Language models are few-shot learners}.
\newblock \bibinfo{journal}{\emph{Advances in neural information processing
  systems}}  \bibinfo{volume}{33} (\bibinfo{year}{2020}),
  \bibinfo{pages}{1877--1901}.
\newblock


\bibitem[Byun et~al\mbox{.}(2023)]%
        {byun2023dispensing}
\bibfield{author}{\bibinfo{person}{Courtni Byun}, \bibinfo{person}{Piper
  Vasicek}, {and} \bibinfo{person}{Kevin Seppi}.}
  \bibinfo{year}{2023}\natexlab{}.
\newblock \showarticletitle{Dispensing with Humans in Human-Computer
  Interaction Research}. In \bibinfo{booktitle}{\emph{Extended Abstracts of the
  2023 CHI Conference on Human Factors in Computing Systems}}.
  \bibinfo{pages}{1--26}.
\newblock


\bibitem[Carrion et~al\mbox{.}(2019)]%
        {carrion2019taxonomy}
\bibfield{author}{\bibinfo{person}{Belen Carrion}, \bibinfo{person}{Teresa
  Onorati}, \bibinfo{person}{Paloma D{\'\i}az}, {and} \bibinfo{person}{Vasiliki
  Triga}.} \bibinfo{year}{2019}\natexlab{}.
\newblock \showarticletitle{A taxonomy generation tool for semantic visual
  analysis of large corpus of documents}.
\newblock \bibinfo{journal}{\emph{Multimedia Tools and Applications}}
  \bibinfo{volume}{78} (\bibinfo{year}{2019}), \bibinfo{pages}{32919--32937}.
\newblock


\bibitem[Chakrabarti et~al\mbox{.}(1997)]%
        {chakrabarti1997using}
\bibfield{author}{\bibinfo{person}{Soumen Chakrabarti}, \bibinfo{person}{Byron
  Dom}, \bibinfo{person}{Rakesh Agrawal}, {and} \bibinfo{person}{Prabhakar
  Raghavan}.} \bibinfo{year}{1997}\natexlab{}.
\newblock \showarticletitle{Using taxonomy, discriminants, and signatures for
  navigating in text databases}. In \bibinfo{booktitle}{\emph{VLDB}},
  Vol.~\bibinfo{volume}{97}. \bibinfo{pages}{446--455}.
\newblock


\bibitem[Chen et~al\mbox{.}(2012)]%
        {chen2012understanding}
\bibfield{author}{\bibinfo{person}{Long Chen}, \bibinfo{person}{Dell Zhang},
  {and} \bibinfo{person}{Levene Mark}.} \bibinfo{year}{2012}\natexlab{}.
\newblock \showarticletitle{Understanding user intent in community question
  answering}. In \bibinfo{booktitle}{\emph{Proceedings of the 21st
  international conference on world wide web}}. \bibinfo{pages}{823--828}.
\newblock


\bibitem[Cheng et~al\mbox{.}(2006)]%
        {cheng2006query}
\bibfield{author}{\bibinfo{person}{Pu-Jeng Cheng},
  \bibinfo{person}{Ching-Hsiang Tsai}, \bibinfo{person}{Chen-Ming Hung}, {and}
  \bibinfo{person}{Lee-Feng Chien}.} \bibinfo{year}{2006}\natexlab{}.
\newblock \showarticletitle{Query taxonomy generation for web search}. In
  \bibinfo{booktitle}{\emph{Proceedings of the 15th ACM international
  conference on Information and knowledge management}}.
  \bibinfo{pages}{862--863}.
\newblock


\bibitem[Chuang and Chien(2003)]%
        {chuang2003automatic}
\bibfield{author}{\bibinfo{person}{Shui-Lung Chuang} {and}
  \bibinfo{person}{Lee-Feng Chien}.} \bibinfo{year}{2003}\natexlab{}.
\newblock \showarticletitle{Automatic query taxonomy generation for information
  retrieval applications}.
\newblock \bibinfo{journal}{\emph{Online Information Review}}
  \bibinfo{volume}{27}, \bibinfo{number}{4} (\bibinfo{year}{2003}),
  \bibinfo{pages}{243--255}.
\newblock


\bibitem[Cohen(1960)]%
        {cohen1960coefficient}
\bibfield{author}{\bibinfo{person}{Jacob Cohen}.}
  \bibinfo{year}{1960}\natexlab{}.
\newblock \showarticletitle{A coefficient of agreement for nominal scales}.
\newblock \bibinfo{journal}{\emph{Educational and psychological measurement}}
  \bibinfo{volume}{20}, \bibinfo{number}{1} (\bibinfo{year}{1960}),
  \bibinfo{pages}{37--46}.
\newblock


\bibitem[Craswell et~al\mbox{.}(2008)]%
        {craswell2008experimental}
\bibfield{author}{\bibinfo{person}{Nick Craswell}, \bibinfo{person}{Onno
  Zoeter}, \bibinfo{person}{Michael Taylor}, {and} \bibinfo{person}{Bill
  Ramsey}.} \bibinfo{year}{2008}\natexlab{}.
\newblock \showarticletitle{An experimental comparison of click position-bias
  models}. In \bibinfo{booktitle}{\emph{Proceedings of the 2008 international
  conference on web search and data mining}}. \bibinfo{pages}{87--94}.
\newblock


\bibitem[Dwivedi et~al\mbox{.}(2023)]%
        {dwivedi2023so}
\bibfield{author}{\bibinfo{person}{Yogesh~K Dwivedi}, \bibinfo{person}{Nir
  Kshetri}, \bibinfo{person}{Laurie Hughes}, \bibinfo{person}{Emma~Louise
  Slade}, \bibinfo{person}{Anand Jeyaraj}, \bibinfo{person}{Arpan~Kumar Kar},
  \bibinfo{person}{Abdullah~M Baabdullah}, \bibinfo{person}{Alex Koohang},
  \bibinfo{person}{Vishnupriya Raghavan}, \bibinfo{person}{Manju Ahuja},
  {et~al\mbox{.}}} \bibinfo{year}{2023}\natexlab{}.
\newblock \showarticletitle{“So what if ChatGPT wrote it?”
  Multidisciplinary perspectives on opportunities, challenges and implications
  of generative conversational AI for research, practice and policy}.
\newblock \bibinfo{journal}{\emph{International Journal of Information
  Management}}  \bibinfo{volume}{71} (\bibinfo{year}{2023}),
  \bibinfo{pages}{102642}.
\newblock


\bibitem[Faggioli et~al\mbox{.}(2023)]%
        {faggioli2023perspectives}
\bibfield{author}{\bibinfo{person}{Guglielmo Faggioli}, \bibinfo{person}{Laura
  Dietz}, \bibinfo{person}{Charles Clarke}, \bibinfo{person}{Gianluca
  Demartini}, \bibinfo{person}{Matthias Hagen}, \bibinfo{person}{Claudia
  Hauff}, \bibinfo{person}{Noriko Kando}, \bibinfo{person}{Evangelos Kanoulas},
  \bibinfo{person}{Martin Potthast}, \bibinfo{person}{Benno Stein},
  {et~al\mbox{.}}} \bibinfo{year}{2023}\natexlab{}.
\newblock \showarticletitle{Perspectives on Large Language Models for Relevance
  Judgment}.
\newblock \bibinfo{journal}{\emph{arXiv preprint arXiv:2304.09161}}
  (\bibinfo{year}{2023}).
\newblock


\bibitem[Fleiss(1971)]%
        {fleiss1971measuring}
\bibfield{author}{\bibinfo{person}{Joseph~L Fleiss}.}
  \bibinfo{year}{1971}\natexlab{}.
\newblock \showarticletitle{Measuring nominal scale agreement among many
  raters.}
\newblock \bibinfo{journal}{\emph{Psychological bulletin}}
  \bibinfo{volume}{76}, \bibinfo{number}{5} (\bibinfo{year}{1971}),
  \bibinfo{pages}{378}.
\newblock


\bibitem[Gao et~al\mbox{.}(2020)]%
        {gao2020recent}
\bibfield{author}{\bibinfo{person}{Jianfeng Gao}, \bibinfo{person}{Chenyan
  Xiong}, {and} \bibinfo{person}{Paul Bennett}.}
  \bibinfo{year}{2020}\natexlab{}.
\newblock \showarticletitle{Recent advances in conversational information
  retrieval}. In \bibinfo{booktitle}{\emph{Proceedings of the 43rd
  International ACM SIGIR Conference on Research and Development in Information
  Retrieval}}. \bibinfo{pages}{2421--2424}.
\newblock


\bibitem[Grandeit et~al\mbox{.}(2020)]%
        {grandeit2020using}
\bibfield{author}{\bibinfo{person}{Philipp Grandeit}, \bibinfo{person}{Carolyn
  Haberkern}, \bibinfo{person}{Maximiliane Lang}, \bibinfo{person}{Jens
  Albrecht}, {and} \bibinfo{person}{Robert Lehmann}.}
  \bibinfo{year}{2020}\natexlab{}.
\newblock \showarticletitle{Using BERT for qualitative content analysis in
  psychosocial online counseling}. In \bibinfo{booktitle}{\emph{Proceedings of
  the Fourth Workshop on Natural Language Processing and Computational Social
  Science}}. \bibinfo{pages}{11--23}.
\newblock


\bibitem[He et~al\mbox{.}(2023)]%
        {he2023annollm}
\bibfield{author}{\bibinfo{person}{Xingwei He}, \bibinfo{person}{Zhenghao Lin},
  \bibinfo{person}{Yeyun Gong}, \bibinfo{person}{Alex Jin},
  \bibinfo{person}{Hang Zhang}, \bibinfo{person}{Chen Lin},
  \bibinfo{person}{Jian Jiao}, \bibinfo{person}{Siu~Ming Yiu},
  \bibinfo{person}{Nan Duan}, \bibinfo{person}{Weizhu Chen}, {et~al\mbox{.}}}
  \bibinfo{year}{2023}\natexlab{}.
\newblock \showarticletitle{Annollm: Making large language models to be better
  crowdsourced annotators}.
\newblock \bibinfo{journal}{\emph{arXiv preprint arXiv:2303.16854}}
  (\bibinfo{year}{2023}).
\newblock


\bibitem[Heer(2019)]%
        {heer2019agency}
\bibfield{author}{\bibinfo{person}{Jeffrey Heer}.}
  \bibinfo{year}{2019}\natexlab{}.
\newblock \showarticletitle{Agency plus automation: Designing artificial
  intelligence into interactive systems}.
\newblock \bibinfo{journal}{\emph{Proceedings of the National Academy of
  Sciences}} \bibinfo{volume}{116}, \bibinfo{number}{6} (\bibinfo{year}{2019}),
  \bibinfo{pages}{1844--1850}.
\newblock


\bibitem[Hope et~al\mbox{.}(2023)]%
        {hope2023computational}
\bibfield{author}{\bibinfo{person}{Tom Hope}, \bibinfo{person}{Doug Downey},
  \bibinfo{person}{Daniel~S Weld}, \bibinfo{person}{Oren Etzioni}, {and}
  \bibinfo{person}{Eric Horvitz}.} \bibinfo{year}{2023}\natexlab{}.
\newblock \showarticletitle{A computational inflection for scientific
  discovery}.
\newblock \bibinfo{journal}{\emph{Commun. ACM}} \bibinfo{volume}{66},
  \bibinfo{number}{8} (\bibinfo{year}{2023}), \bibinfo{pages}{62--73}.
\newblock


\bibitem[Jansen et~al\mbox{.}(2007)]%
        {jansen2007determining}
\bibfield{author}{\bibinfo{person}{Bernard~J Jansen},
  \bibinfo{person}{Danielle~L Booth}, {and} \bibinfo{person}{Amanda Spink}.}
  \bibinfo{year}{2007}\natexlab{}.
\newblock \showarticletitle{Determining the user intent of web search engine
  queries}. In \bibinfo{booktitle}{\emph{Proceedings of the 16th international
  conference on World Wide Web}}. \bibinfo{pages}{1149--1150}.
\newblock


\bibitem[Jeronymo et~al\mbox{.}(2023)]%
        {jeronymo2023inpars}
\bibfield{author}{\bibinfo{person}{Vitor Jeronymo}, \bibinfo{person}{Luiz
  Bonifacio}, \bibinfo{person}{Hugo Abonizio}, \bibinfo{person}{Marzieh
  Fadaee}, \bibinfo{person}{Roberto Lotufo}, \bibinfo{person}{Jakub Zavrel},
  {and} \bibinfo{person}{Rodrigo Nogueira}.} \bibinfo{year}{2023}\natexlab{}.
\newblock \showarticletitle{InPars-v2: Large Language Models as Efficient
  Dataset Generators for Information Retrieval}.
\newblock \bibinfo{journal}{\emph{arXiv preprint arXiv:2301.01820}}
  (\bibinfo{year}{2023}).
\newblock


\bibitem[Kaplan et~al\mbox{.}(2022)]%
        {kaplan2022introducing}
\bibfield{author}{\bibinfo{person}{Angelika Kaplan}, \bibinfo{person}{Thomas
  K{\"u}hn}, \bibinfo{person}{Sebastian Hahner}, \bibinfo{person}{Niko
  Benkler}, \bibinfo{person}{Jan Keim}, \bibinfo{person}{Dominik Fuch{\ss}},
  \bibinfo{person}{Sophie Corallo}, {and} \bibinfo{person}{Robert Heinrich}.}
  \bibinfo{year}{2022}\natexlab{}.
\newblock \showarticletitle{Introducing an Evaluation Method for Taxonomies}.
  In \bibinfo{booktitle}{\emph{Proceedings of the 26th International Conference
  on Evaluation and Assessment in Software Engineering}}.
  \bibinfo{pages}{311--316}.
\newblock


\bibitem[Kellar et~al\mbox{.}(2007)]%
        {kellar2007field}
\bibfield{author}{\bibinfo{person}{Melanie Kellar}, \bibinfo{person}{Carolyn
  Watters}, {and} \bibinfo{person}{Michael Shepherd}.}
  \bibinfo{year}{2007}\natexlab{}.
\newblock \showarticletitle{A field study characterizing Web-based
  information-seeking tasks}.
\newblock \bibinfo{journal}{\emph{Journal of the American Society for
  information science and technology}} \bibinfo{volume}{58},
  \bibinfo{number}{7} (\bibinfo{year}{2007}), \bibinfo{pages}{999--1018}.
\newblock


\bibitem[Kong et~al\mbox{.}(2015)]%
        {kong2015predicting}
\bibfield{author}{\bibinfo{person}{Weize Kong}, \bibinfo{person}{Rui Li},
  \bibinfo{person}{Jie Luo}, \bibinfo{person}{Aston Zhang}, \bibinfo{person}{Yi
  Chang}, {and} \bibinfo{person}{James Allan}.}
  \bibinfo{year}{2015}\natexlab{}.
\newblock \showarticletitle{Predicting search intent based on pre-search
  context}. In \bibinfo{booktitle}{\emph{Proceedings of the 38th International
  ACM SIGIR Conference on Research and Development in Information Retrieval}}.
  \bibinfo{pages}{503--512}.
\newblock


\bibitem[Kundisch et~al\mbox{.}(2021)]%
        {kundisch2021update}
\bibfield{author}{\bibinfo{person}{Dennis Kundisch}, \bibinfo{person}{Jan
  Muntermann}, \bibinfo{person}{Anna~Maria Oberl{\"a}nder},
  \bibinfo{person}{Daniel Rau}, \bibinfo{person}{Maximilian R{\"o}glinger},
  \bibinfo{person}{Thorsten Schoormann}, {and} \bibinfo{person}{Daniel
  Szopinski}.} \bibinfo{year}{2021}\natexlab{}.
\newblock \showarticletitle{An update for taxonomy designers: methodological
  guidance from information systems research}.
\newblock \bibinfo{journal}{\emph{Business \& Information Systems Engineering}}
  (\bibinfo{year}{2021}), \bibinfo{pages}{1--19}.
\newblock


\bibitem[Li et~al\mbox{.}(2022)]%
        {li2022user}
\bibfield{author}{\bibinfo{person}{Shuokai Li}, \bibinfo{person}{Ruobing Xie},
  \bibinfo{person}{Yongchun Zhu}, \bibinfo{person}{Xiang Ao},
  \bibinfo{person}{Fuzhen Zhuang}, {and} \bibinfo{person}{Qing He}.}
  \bibinfo{year}{2022}\natexlab{}.
\newblock \showarticletitle{User-centric conversational recommendation with
  multi-aspect user modeling}. In \bibinfo{booktitle}{\emph{Proceedings of the
  45th International ACM SIGIR Conference on Research and Development in
  Information Retrieval}}. \bibinfo{pages}{223--233}.
\newblock


\bibitem[Li and Belkin(2008)]%
        {li2008faceted}
\bibfield{author}{\bibinfo{person}{Yuelin Li} {and} \bibinfo{person}{Nicholas~J
  Belkin}.} \bibinfo{year}{2008}\natexlab{}.
\newblock \showarticletitle{A faceted approach to conceptualizing tasks in
  information seeking}.
\newblock \bibinfo{journal}{\emph{Information processing \& management}}
  \bibinfo{volume}{44}, \bibinfo{number}{6} (\bibinfo{year}{2008}),
  \bibinfo{pages}{1822--1837}.
\newblock


\bibitem[Lippell(2022)]%
        {lippell2022taxonomies}
\bibfield{author}{\bibinfo{person}{Helen Lippell}.}
  \bibinfo{year}{2022}\natexlab{}.
\newblock \bibinfo{booktitle}{\emph{Taxonomies: Practical Approaches to
  Developing and Managing Vocabularies for Digital Information}}.
\newblock \bibinfo{publisher}{Facet Publishing}.
\newblock


\bibitem[Liu et~al\mbox{.}(2019)]%
        {liu2019task}
\bibfield{author}{\bibinfo{person}{Jiqun Liu}, \bibinfo{person}{Matthew
  Mitsui}, \bibinfo{person}{Nicholas~J Belkin}, {and} \bibinfo{person}{Chirag
  Shah}.} \bibinfo{year}{2019}\natexlab{}.
\newblock \showarticletitle{Task, information seeking intentions, and user
  behavior: Toward a multi-level understanding of Web search}. In
  \bibinfo{booktitle}{\emph{Proceedings of the 2019 conference on human
  information interaction and retrieval}}. \bibinfo{pages}{123--132}.
\newblock


\bibitem[Mitsui et~al\mbox{.}(2016)]%
        {mitsui2016extracting}
\bibfield{author}{\bibinfo{person}{Matthew Mitsui}, \bibinfo{person}{Chirag
  Shah}, {and} \bibinfo{person}{Nicholas~J Belkin}.}
  \bibinfo{year}{2016}\natexlab{}.
\newblock \showarticletitle{Extracting information seeking intentions for web
  search sessions}. In \bibinfo{booktitle}{\emph{Proceedings of the 39th
  International ACM SIGIR conference on Research and Development in Information
  Retrieval}}. \bibinfo{pages}{841--844}.
\newblock


\bibitem[M{\"o}kander et~al\mbox{.}(2023)]%
        {mokander2023auditing}
\bibfield{author}{\bibinfo{person}{Jakob M{\"o}kander}, \bibinfo{person}{Jonas
  Schuett}, \bibinfo{person}{Hannah~Rose Kirk}, {and} \bibinfo{person}{Luciano
  Floridi}.} \bibinfo{year}{2023}\natexlab{}.
\newblock \showarticletitle{Auditing large language models: a three-layered
  approach}.
\newblock \bibinfo{journal}{\emph{AI and Ethics}} (\bibinfo{year}{2023}),
  \bibinfo{pages}{1--31}.
\newblock


\bibitem[Moore et~al\mbox{.}(2023)]%
        {moore2023empowering}
\bibfield{author}{\bibinfo{person}{Steven Moore}, \bibinfo{person}{Richard
  Tong}, \bibinfo{person}{Anjali Singh}, \bibinfo{person}{Zitao Liu},
  \bibinfo{person}{Xiangen Hu}, \bibinfo{person}{Yu Lu},
  \bibinfo{person}{Joleen Liang}, \bibinfo{person}{Chen Cao},
  \bibinfo{person}{Hassan Khosravi}, \bibinfo{person}{Paul Denny},
  {et~al\mbox{.}}} \bibinfo{year}{2023}\natexlab{}.
\newblock \showarticletitle{Empowering Education with LLMs-The Next-Gen
  Interface and Content Generation}. In \bibinfo{booktitle}{\emph{International
  Conference on Artificial Intelligence in Education}}. Springer,
  \bibinfo{pages}{32--37}.
\newblock


\bibitem[Morzy et~al\mbox{.}(2000)]%
        {morzy2000data}
\bibfield{author}{\bibinfo{person}{Tadeusz Morzy}, \bibinfo{person}{Marek
  Wojciechowski}, {and} \bibinfo{person}{Maciej Zakrzewicz}.}
  \bibinfo{year}{2000}\natexlab{}.
\newblock \showarticletitle{Data mining support in database management
  systems}. In \bibinfo{booktitle}{\emph{Data Warehousing and Knowledge
  Discovery: Second International Conference, DaWaK 2000 London, UK, September
  4--6, 2000 Proceedings 2}}. Springer, \bibinfo{pages}{382--392}.
\newblock


\bibitem[Raad and Cruz(2015)]%
        {raad2015survey}
\bibfield{author}{\bibinfo{person}{Joe Raad} {and} \bibinfo{person}{Christophe
  Cruz}.} \bibinfo{year}{2015}\natexlab{}.
\newblock \showarticletitle{A survey on ontology evaluation methods}. In
  \bibinfo{booktitle}{\emph{Proceedings of the International Conference on
  Knowledge Engineering and Ontology Development, part of the 7th International
  Joint Conference on Knowledge Discovery, Knowledge Engineering and Knowledge
  Management}}.
\newblock


\bibitem[Radlinski and Craswell(2017)]%
        {radlinski2017theoretical}
\bibfield{author}{\bibinfo{person}{Filip Radlinski} {and} \bibinfo{person}{Nick
  Craswell}.} \bibinfo{year}{2017}\natexlab{}.
\newblock \showarticletitle{A theoretical framework for conversational search}.
  In \bibinfo{booktitle}{\emph{Proceedings of the 2017 conference on conference
  human information interaction and retrieval}}. \bibinfo{pages}{117--126}.
\newblock


\bibitem[Ren et~al\mbox{.}(2021)]%
        {ren2021wizard}
\bibfield{author}{\bibinfo{person}{Pengjie Ren}, \bibinfo{person}{Zhongkun
  Liu}, \bibinfo{person}{Xiaomeng Song}, \bibinfo{person}{Hongtao Tian},
  \bibinfo{person}{Zhumin Chen}, \bibinfo{person}{Zhaochun Ren}, {and}
  \bibinfo{person}{Maarten de Rijke}.} \bibinfo{year}{2021}\natexlab{}.
\newblock \showarticletitle{Wizard of search engine: Access to information
  through conversations with search engines}. In
  \bibinfo{booktitle}{\emph{Proceedings of the 44th International ACM SIGIR
  Conference on research and development in information retrieval}}.
  \bibinfo{pages}{533--543}.
\newblock


\bibitem[Rose and Levinson(2004)]%
        {rose2004understanding}
\bibfield{author}{\bibinfo{person}{Daniel~E Rose} {and} \bibinfo{person}{Danny
  Levinson}.} \bibinfo{year}{2004}\natexlab{}.
\newblock \showarticletitle{Understanding user goals in web search}. In
  \bibinfo{booktitle}{\emph{Proceedings of the 13th international conference on
  World Wide Web}}. \bibinfo{pages}{13--19}.
\newblock


\bibitem[Singhal et~al\mbox{.}(2023)]%
        {singhal2023large}
\bibfield{author}{\bibinfo{person}{Karan Singhal}, \bibinfo{person}{Shekoofeh
  Azizi}, \bibinfo{person}{Tao Tu}, \bibinfo{person}{S~Sara Mahdavi},
  \bibinfo{person}{Jason Wei}, \bibinfo{person}{Hyung~Won Chung},
  \bibinfo{person}{Nathan Scales}, \bibinfo{person}{Ajay Tanwani},
  \bibinfo{person}{Heather Cole-Lewis}, \bibinfo{person}{Stephen Pfohl},
  {et~al\mbox{.}}} \bibinfo{year}{2023}\natexlab{}.
\newblock \showarticletitle{Large language models encode clinical knowledge}.
\newblock \bibinfo{journal}{\emph{Nature}} (\bibinfo{year}{2023}),
  \bibinfo{pages}{1--9}.
\newblock


\bibitem[Spangler and Kreulen(2002)]%
        {spangler2002interactive}
\bibfield{author}{\bibinfo{person}{Scott Spangler} {and}
  \bibinfo{person}{Jeffrey Kreulen}.} \bibinfo{year}{2002}\natexlab{}.
\newblock \showarticletitle{Interactive methods for taxonomy editing and
  validation}. In \bibinfo{booktitle}{\emph{Proceedings of the eleventh
  international conference on Information and knowledge management}}.
  \bibinfo{pages}{665--668}.
\newblock


\bibitem[Teevan et~al\mbox{.}(2007)]%
        {teevan2007information}
\bibfield{author}{\bibinfo{person}{Jaime Teevan}, \bibinfo{person}{Eytan Adar},
  \bibinfo{person}{Rosie Jones}, {and} \bibinfo{person}{Michael~AS Potts}.}
  \bibinfo{year}{2007}\natexlab{}.
\newblock \showarticletitle{Information re-retrieval: Repeat queries in Yahoo's
  logs}. In \bibinfo{booktitle}{\emph{Proceedings of the 30th annual
  international ACM SIGIR conference on Research and development in information
  retrieval}}. \bibinfo{pages}{151--158}.
\newblock


\bibitem[T{\"o}rnberg(2023)]%
        {tornberg2023use}
\bibfield{author}{\bibinfo{person}{Petter T{\"o}rnberg}.}
  \bibinfo{year}{2023}\natexlab{}.
\newblock \showarticletitle{How to use LLMs for Text Analysis}.
\newblock \bibinfo{journal}{\emph{arXiv preprint arXiv:2307.13106}}
  (\bibinfo{year}{2023}).
\newblock


\bibitem[Vakkari(2016)]%
        {vakkari2016searching}
\bibfield{author}{\bibinfo{person}{Pertti Vakkari}.}
  \bibinfo{year}{2016}\natexlab{}.
\newblock \showarticletitle{Searching as learning: A systematization based on
  literature}.
\newblock \bibinfo{journal}{\emph{Journal of Information Science}}
  \bibinfo{volume}{42}, \bibinfo{number}{1} (\bibinfo{year}{2016}),
  \bibinfo{pages}{7--18}.
\newblock


\bibitem[Vtyurina et~al\mbox{.}(2020)]%
        {vtyurina2020mixed}
\bibfield{author}{\bibinfo{person}{Alexandra Vtyurina},
  \bibinfo{person}{Charles~LA Clarke}, \bibinfo{person}{Edith Law},
  \bibinfo{person}{Johanne~R Trippas}, {and} \bibinfo{person}{Horatiu Bota}.}
  \bibinfo{year}{2020}\natexlab{}.
\newblock \showarticletitle{A mixed-method analysis of text and audio search
  interfaces with varying task complexity}. In
  \bibinfo{booktitle}{\emph{Proceedings of the 2020 ACM SIGIR on International
  Conference on Theory of Information Retrieval}}. \bibinfo{pages}{61--68}.
\newblock


\bibitem[Watkins(2023)]%
        {watkins2023guidance}
\bibfield{author}{\bibinfo{person}{Ryan Watkins}.}
  \bibinfo{year}{2023}\natexlab{}.
\newblock \showarticletitle{Guidance for researchers and peer-reviewers on the
  ethical use of Large Language Models (LLMs) in scientific research
  workflows}.
\newblock \bibinfo{journal}{\emph{AI and Ethics}} (\bibinfo{year}{2023}),
  \bibinfo{pages}{1--6}.
\newblock


\bibitem[White and Drucker(2007)]%
        {white2007investigating}
\bibfield{author}{\bibinfo{person}{Ryen~W White} {and}
  \bibinfo{person}{Steven~M Drucker}.} \bibinfo{year}{2007}\natexlab{}.
\newblock \showarticletitle{Investigating behavioral variability in web
  search}. In \bibinfo{booktitle}{\emph{Proceedings of the 16th international
  conference on World Wide Web}}. \bibinfo{pages}{21--30}.
\newblock


\bibitem[Xie(2002)]%
        {xie2002patterns}
\bibfield{author}{\bibinfo{person}{Hong Xie}.} \bibinfo{year}{2002}\natexlab{}.
\newblock \showarticletitle{Patterns between interactive intentions and
  information-seeking strategies}.
\newblock \bibinfo{journal}{\emph{Information processing and Management}}
  \bibinfo{volume}{38}, \bibinfo{number}{1} (\bibinfo{year}{2002}),
  \bibinfo{pages}{55--77}.
\newblock


\bibitem[Yang(2012)]%
        {yang2012constructing}
\bibfield{author}{\bibinfo{person}{Hui Yang}.} \bibinfo{year}{2012}\natexlab{}.
\newblock \showarticletitle{Constructing task-specific taxonomies for document
  collection browsing}. In \bibinfo{booktitle}{\emph{Proceedings of the 2012
  Joint Conference on Empirical Methods in Natural Language Processing and
  Computational Natural Language Learning}}. \bibinfo{pages}{1278--1289}.
\newblock


\bibitem[Zamir and Etzioni(1998)]%
        {zamir1998web}
\bibfield{author}{\bibinfo{person}{Oren Zamir} {and} \bibinfo{person}{Oren
  Etzioni}.} \bibinfo{year}{1998}\natexlab{}.
\newblock \showarticletitle{Web document clustering: A feasibility
  demonstration}. In \bibinfo{booktitle}{\emph{Proceedings of the 21st annual
  international ACM SIGIR conference on Research and development in information
  retrieval}}. \bibinfo{pages}{46--54}.
\newblock


\bibitem[Zhang et~al\mbox{.}(2019)]%
        {zhang2019generic}
\bibfield{author}{\bibinfo{person}{Hongfei Zhang}, \bibinfo{person}{Xia Song},
  \bibinfo{person}{Chenyan Xiong}, \bibinfo{person}{Corby Rosset},
  \bibinfo{person}{Paul~N Bennett}, \bibinfo{person}{Nick Craswell}, {and}
  \bibinfo{person}{Saurabh Tiwary}.} \bibinfo{year}{2019}\natexlab{}.
\newblock \showarticletitle{Generic intent representation in web search}. In
  \bibinfo{booktitle}{\emph{Proceedings of the 42nd International ACM SIGIR
  Conference on Research and Development in Information Retrieval}}.
  \bibinfo{pages}{65--74}.
\newblock


\bibitem[Zhang et~al\mbox{.}(2023)]%
        {zhang2023efficiently}
\bibfield{author}{\bibinfo{person}{Peiyan Zhang}, \bibinfo{person}{Jiayan Guo},
  \bibinfo{person}{Chaozhuo Li}, \bibinfo{person}{Yueqi Xie},
  \bibinfo{person}{Jae~Boum Kim}, \bibinfo{person}{Yan Zhang},
  \bibinfo{person}{Xing Xie}, \bibinfo{person}{Haohan Wang}, {and}
  \bibinfo{person}{Sunghun Kim}.} \bibinfo{year}{2023}\natexlab{}.
\newblock \showarticletitle{Efficiently leveraging multi-level user intent for
  session-based recommendation via atten-mixer network}. In
  \bibinfo{booktitle}{\emph{Proceedings of the Sixteenth ACM International
  Conference on Web Search and Data Mining}}. \bibinfo{pages}{168--176}.
\newblock


\end{thebibliography}

\vspace*{-0.5em}
\section{Ethical Considerations}
The work described here heavily relies on LLMs, which are shown to have several issues in their training (e.g., bias in datasets used), application (e.g., toxicity and hallucination), as well as user perceptions (blindly believing all responses due to implicit trust in such systems). Our work is not associated with direct use of LLMs for end-users, but it does concern itself with interpreting what those users may be doing using LLMs. The main contribution in this paper is a methodology that can be used for developing and using user intent taxonomies. Given the efficiency and effectiveness of this method, one may be inclined to use it in a wide range of applications that may not be advisable. For example, this method should not be used on certain vulnerable populations such as children and people with different disabilities. The results -- either in terms of constructing taxonomies to understand their behavior or in terms of applying taxonomies to generate insights or make recommendations -- could be misleading and potentially harmful. We note that while we were using data that is not publicly available, it lacked any identifying information about the users. Abundance of caution was taken to ensure privacy and protection of this data and the users. However, someone else using our method on other datasets must do their own due diligence to protect the users from any potential harmful effects.

\end{document}